# Coordinated Analyses of Presolar Grains in the Allan Hills 77307 and Queen Elizabeth Range 99177 Meteorites

Short Title: Presolar grains in ALHA 77307 and QUE 99177

Ann N. Nguyen[1,*], Larry R. Nittler[1], Frank J. Stadermann[2], Rhonda M. Stroud[3], and Conel M. O'D. Alexander[1]

[1] Department of Terrestrial Magnetism, Carnegie Institution of Washington, 5241 Broad Branch Road, NW, Washington, DC 20015. [2] Laboratory for Space Sciences and the Physics Department, Washington University, One Brookings Drive, St. Louis, MO 63130. [3] Naval Research Laboratory, Code 6366, Washington, DC 20375.

* Corresponding Author: Robert M. Walker Laboratory for Space Science, Astromaterials Research and Exploration Science Directorate, NASA Johnson Space Center, Houston, TX 77058; ESC Group/Jacobs Sverdrup, NASA Johnson Space Center, Houston, TX 77058. Email address: lan-anh.n.nguyen@nasa.gov

ABSTRACT

We report the identification of presolar silicates (~177 ppm), presolar oxides (~11 ppm), and one presolar $SiO_2$ grain in the Allan Hills (ALHA) 77307 chondrite. Three grains having Si isotopic compositions similar to SiC X and Z grains were also identified, though the mineral phases are unconfirmed. Similar abundances of presolar silicates (~152 ppm) and oxides (~8 ppm) were also uncovered in the primitive CR chondrite Queen Elizabeth Range (QUE) 99177, along with 13 presolar SiC grains and one presolar silicon nitride. The O isotopic compositions of the presolar silicates and oxides indicate that most of the grains condensed in low-mass red giant and asymptotic giant branch stars. Interestingly, unlike presolar oxides, few presolar silicate grains have isotopic compositions pointing to low-metallicity, low-mass stars (Group 3). The $^{18}O$-rich (Group 4) silicates, along with the few Group 3 silicates that were identified, likely have origins in supernova outflows. This is supported by their O and Si isotopic compositions.

Elemental compositions for 74 presolar silicate grains were determined by scanning Auger spectroscopy. Most of the grains have non-stoichiometric elemental compositions inconsistent with pyroxene or olivine, the phases commonly used to fit astronomical spectra, and have comparable Mg and Fe contents. Non-equilibrium condensation and/or secondary alteration could produce the high Fe contents. Transmission electron microscopic analysis of three silicate grains also reveals non-stoichiometric compositions, attributable to non-equilibrium or multistep condensation, and very fine scale elemental heterogeneity, possibly due to subsequent annealing. The mineralogies of presolar silicates identified in meteorites thus far seem to differ from those in interplanetary dust particles.

## 1. INTRODUCTION

Dust grains that condensed in the atmospheres of evolved stars and in supernova (SN) ejecta were transported through the interstellar medium (ISM) and incorporated into our forming solar system 4.6 Gyr ago. Despite this long history, these "presolar" grains retain the isotopic compositions of their parent sources and reflect various astrophysical processes. Meteorites, interplanetary dust particles (IDPs), Antarctic micrometeorites, and cometary dust harbor presolar grains (e.g. Zinner 2007), which are identified in the laboratory by their distinctive isotopic signatures. Insight into stellar evolution, nuclear processes, and Galactic chemical evolution can be gleaned from the isotopic compositions of these grains. The chemical compositions and mineralogies of presolar grains can often be identified as well and provide further detail about dust condensation and modification processes.

The study of presolar grains is a relatively recent endeavor, with the first discovery made on the basis of exotic noble gas signatures in presolar diamond, SiC, and graphite (Amari et al. 1990; Bernatowicz et al. 1987; Lewis et al. 1987). However, O-rich dust comprises the bulk of dust observed around evolved stars and in the ISM. Thus, a major piece of the circumstellar picture in the form of O-rich presolar grains was initially missing. Carbonaceous phases can be chemically isolated and do not suffer from a background of solar system C-rich minerals. On the other hand, the main components of meteorites are oxide and silicate grains that formed in the solar system. Thus, the identification of O-rich presolar grains necessitates the isotopic analysis of a large number of grains. For many oxide species (e.g., $Al_2O_3$), this can be aided by ion microprobe studies of acid residues in which the dominant O-bearing phases (silicates) have been destroyed. Indeed, several presolar oxide phases have been identified in meteoritic acid residues, including corundum ($Al_2O_3$), spinel ($MgAl_2O_4$), hibonite ($CaAl_{12}O_{19}$), chromite ($(Fe, Mg)Cr_2O_4$), and $TiO_2$.

Isotopically anomalous silicates clearly cannot be identified in acid residues, and their analysis is thus restricted to chemically untreated samples. Moreover, along with the large background of solar system silicates, the identification of presolar silicate grains is further complicated by their submicron sizes. The first discovery of presolar silicates was made in IDPs using the exceptional capabilities of the Cameca NanoSIMS 50 ion microprobe (Messenger et al. 2003a). Specifically, this instrument achieves high sensitivity at high (submicron) lateral resolution. Shortly after, silicate stardust grains were identified in meteorites using both the NanoSIMS and Cameca ims-1270 ion microprobes (Mostefaoui & Hoppe 2004; Nagashima, Krot, & Yurimoto 2004; Nguyen & Zinner 2004).

The oxygen isotopic compositions of presolar silicate (and oxide) grains agree with observations and astrophysical models of red giant branch (RGB) and asymptotic giant branch (AGB) stars and SN (Choi et al. 1998; Nittler et al. 1997b; Nittler et al. 1998). These isotopic compositions are not only used to identify the parent sources, they also help constrain models of stellar evolution, convective mixing processes (where material from inner stellar regions is mixed into the envelope), nucleosynthesis and GCE. In addition, the compositions and physical characteristics of the parent stellar atmospheres can be studied by determining the chemical compositions of these grains. Further information about dust condensation can be gleaned from the crystal structure (or lack thereof) of stardust grains.

Silicate grains are the most abundant condensates around O-rich evolved stars (Demyk et al. 2000; Waters et al. 1996). Yet only a few hundred presolar silicates have been identified in extraterrestrial materials to date, compared to over 600 and 8000 presolar oxides and SiC, respectively (Hynes & Gyngard 2009). Clearly the characterization of these circumstellar silicate grains is still in its infancy. By applying novel complementary techniques that operate on the sub-micrometer scale, we have made strides to extract as much information as possible out of the grains identified in this study, and thus learn about their parent stars. The oxygen and silicon isotopic compositions of presolar phases were identified *in situ* in the highly primitive carbonaceous chondrites Allan Hills (ALHA) 77307 (CO3) and Queen Elizabeth Range (QUE) 99177 (CR2) by the Carnegie NanoSIMS 50L. The mineralogies and chemical compositions of three grains in cross-section were obtained by transmission electron microscopy (TEM). Additionally, Auger microscopy was applied to determine the chemical compositions of many of the presolar grains identified in ALHA 77307. We detail these experimental techniques and discuss the astrophysical relevance of our results.

## 2. EXPERIMENTAL

We chose matrix areas of polished thin sections of the ALHA 77307 and QUE 99177 carbonaceous chondrites for analysis with the Carnegie NanoSIMS 50L. ALHA 77307 has already been shown to contain a high abundance of presolar silicates (Kobayashi et al. 2005; Nguyen et al. 2007b). QUE 99177 (and MET 00426), unlike most other CR chondrites, has recently been shown to have experienced only mild hydration (Abreu & Brearley 2010) and also to contain abundant presolar silicates (Floss & Stadermann 2009).

Compared to the previous-generation NanoSIMS 50 used in most other studies of presolar silicates in meteorites (Floss & Stadermann 2009; Mostefaoui & Hoppe 2004; Nguyen et al. 2007b; Nguyen & Zinner 2004; Vollmer, Hoppe, & Brenker 2008) the mass spectrometer of the NanoSIMS 50L has a larger electromagnet and two additional moveable ion detectors allowing more flexibility in measurements. For the present work, a focused $Cs^+$ primary ion beam of ~1pA was rastered over 20 × 20 $\mu m^2$ areas for 10-30 scans of 256 × 256 pixels each. Under these conditions, the nominal primary beam diameter (and thus ideal spatial resolution) is ~100 nm. However, it is in general very difficult to determine the exact size and shape of the primary beam. Moreover, although image shifts between scans (typically only a few pixels) are corrected for, such shifts are often non-uniform across an image and the precision with which they can be determined is limited. As a result, the effective spatial resolution is in most cases probably larger than the true beam size, which may be larger than 100 nm and asymmetric due to slight tuning changes across a sample. This is true of all NanoSIMS imaging searches for presolar grains.

The total analyzed matrix areas were determined by setting a threshold on the summed $^{16}O^-$ image of each analysis region. Total matrix areas of 42880 $\mu m^2$ and 21170 $\mu m^2$ were analyzed in ALHA 77307 and QUE 99177, respectively. For most of the ALHA 77307 measurements (19850 $\mu m^2$), the three O isotopes, three Si isotopes, and $^{24}Mg^{16}O$ were measured simultaneously as negative secondary ions in seven electron multipliers, along with secondary electrons. For some regions (12910 $\mu m^2$), $^{27}Al^{16}O$ was measured rather than $^{24}Mg^{16}O$. Silicon-29 and $^{30}Si$ were not measured for an area of 7540 $\mu m^2$. For a smaller area (1250 $\mu m^2$) of ALHA 77307 and for all measurements of QUE 99177, $^{12}C$, $^{13}C$, the three O isotopes, $^{28}Si$, $^{30}Si$ (the latter only in QUE 99177) and secondary electrons were measured simultaneously. Many of these same areas were subsequently analyzed for N and/or H isotopic compositions. For the most part, the C, N and H measurements targeted organic matter, not presolar stardust grains, and these data will be presented and discussed elsewhere (Alexander et al. in preparation).

For all presolar grain searches, the mass resolution was adequate to ensure separation of isobaric interferences. The number of scans was adjusted so that the 1σ counting statistical error for $^{17}O/^{16}O$ in 250 nm silicate grains was no more than 15%, and typically less than 10%. Measurement times for a 20 × 20 μm$^2$ area ranged from about 2 to 5 hours. The resulting 256 × 256 pixel ion images were processed using the L'Image custom software (L. R. Nittler, Carnegie Institution). Any shifts between scans are corrected for, and isotopic ratio images are produced. From these integrated ratio images, grains having anomalous O or Si isotopic ratios relative to the surrounding matrix material can generally be identified and regions of interest were manually defined to obtain their isotopic compositions. Moreover, each image was sub-divided into 3×3 pixel (~235 nm wide) regions to determine whether any apparent anomalies were indeed statistically significant and also to uncover any overlooked anomalous regions. The primary component of carbonaceous chondrites is silicate grains of solar system origin. We thus used this material as internal isotopic standards, and all reported O and Si ratios for silicates and oxides are normalized to these "normal" isotopic compositions. Measurement of synthetic SiC standards revealed a ~12‰ amu$^{-1}$ instrumental fractionation for Si isotopes in SiC, relative to the bulk matrix of QUE 99177. The Si data for SiC grains in this meteorite were thus corrected for this fractionation. Note that the matrices of carbonaceous chondrites have O-isotopic compositions within 15‰ of terrestrial values and essentially terrestrial Si-isotopic compositions. Since the measurement errors (from counting statistics) for individual sub-micrometer grains in our images are much larger than this, this internal normalization procedure does not introduce significant additional uncertainty. Initial silicate and oxide phase distinctions for presolar grains were made based upon the measured $^{28}Si^-/^{16}O^-$ ratios, relative to the normal matrix grains.

Several consequences of raster ion imaging of thin sections and densely packed grain dispersions have previously been discussed (Nguyen, Zinner, & Lewis 2003; Nguyen et al. 2007b). In particular, isotopic dilution caused by primary beam overlap onto normal grains shifts any anomalous isotopic ratios toward the solar composition. Analysis of simulated images generated with the assumption of a Gaussian primary beam density profile (Nguyen et al. 2007b) indicates that even for grains nominally a few times larger than the primary beam diameter, a significant proportion of the measured signal is from neighboring material. Thus, the actual isotopic ratios are in most (and probably all) cases more extreme than those reported. Moreover, grains having marginal anomalies are almost certainly missed by ion imaging and the reported presolar grain abundances are hence lower limits. These dilution effects are also a concern for single grain measurements if there is overlap of the primary beam onto solar system grains. This topic will be discussed in more detail in a later section. Phase designations (e.g., oxide versus silicate) based purely on the NanoSIMS $^{28}Si^-/^{16}O^-$ ratios should also be considered preliminary because of this beam overlap. Note that this effect is size-dependent and affects smaller grains much more than the comparatively larger ones. As discussed below, and as also found by Nguyen et al. (2007b), in some cases the NanoSIMS classification was indeed found to be incorrect based on subsequent Auger analyses.

The magnitudes of Si isotopic anomalies in most presolar grains are much smaller than those of O isotopic anomalies (Amari, Zinner, & Lewis 1995; Hoppe et al. 1995; Hoppe et al. 1993; Mostefaoui & Hoppe 2004; Nittler et al. 1997b; Vollmer et al. 2008). Thus, extraction of useful astrophysical information from Si isotopes requires higher precision than for O isotopes.

Although the greater number of detectors on the NanoSIMS 50L allowed us to measure Si isotopic compositions along with O isotopes, for the most part the measurements resulted in similar counting-statistical precisions for Si and O isotopes (the relatively higher abundance of the rare Si isotopes compared to O isotopes is counter-balanced by the lower negative secondary ion yield of Si in silicates). Higher-precision Si isotopic data were acquired by re-measurement of a few presolar grains in each meteorite, chosen either because they were larger than average or had unusual O isotopic compositions. The effect of isotope dilution from neighboring material on the measured Si isotope compositions is even more severe for Si than for O since the anticipated isotopic effects are smaller. This important issue is discussed in detail in section 4.2.

Identified presolar silicate grains have typical diameters of only ~300 nm and their chemical compositions cannot be determined *in situ* by energy dispersive X-ray spectroscopy (EDX) in a secondary electron microscope (SEM) due to the relatively large X-ray excitation volume. However, with a lateral resolution and excitation volume in the tens of nanometers, scanning Auger spectroscopy is capable of spatially resolving individual presolar silicate grains and determining their major element compositions (Stadermann et al. 2009). In fact, in both Nguyen et al. (2007b) and the present work, some initial grain classifications based on NanoSIMS secondary ion yields had to be corrected when this additional data became available. The Auger measurements were performed with a PHI 700 Scanning Auger Nanoprobe at Washington University, St. Louis, largely following the routine outlined in Stadermann et al. (2009). Prior to analysis, the sample is briefly sputtered with a low-energy, wide-area Ar beam to remove surface contaminants. Individual grains are measured by manually defining regions of interest as rectangles from secondary electron (SE) images and by rastering a ~10 kV primary beam of 0.25 nA for about 30 minutes over these regions, while acquiring a series of Auger electron energy spectra in the range from 30 to 1730 eV with step sizes of 1 eV. These spectra are then averaged, smoothed and differentiated using a seven-point Savitzky-Golay algorithm. Relative elemental quantification (normalized to 100 at. %) of these results is based on peak heights in the derivative spectra and on sensitivity factors obtained from measurement of silicate standards (Stadermann et al. 2009). In some cases, grain boundaries were not clear in SE images and the placement of Auger analysis regions were subsequently verified.

In addition to acquiring compositional data on individual grains, Auger spectroscopy was also used to obtain elemental distribution images at high spatial resolution. For these measurements a primary beam of 5 - 10 nA is used to produce 256 × 256 pixel maps of 5 × 5 $\mu m^2$ areas of the major elements O, Si, Mg, Fe, Al and Ca. The acquisition of such maps can take up to several hours for each element. Grain compositions cannot be quantified from these maps, but they are often useful in showing grain boundaries, which may not be visible in SE images, and possible elemental heterogeneity. For both measurement types, image drift is corrected for by automated image registration. In addition, SE images of the region surrounding the analyzed grain taken before and after the measurement are compared to ensure that there was no drift. This was done for about half of the Auger measurements. It is important to note that Auger spectroscopy is a surface analytical technique which measures the composition of the top few nanometers of the sample. It does not provide information on the depth (thickness) of a given grain or on elemental heterogeneities in the cross-sectional direction.

While Auger spectroscopy obtains the near-surface chemical compositions only, analysis of extracted grain cross-sections by TEM can provide definitive mineral (structural and chemical) classifications of grains not completely consumed by NanoSIMS analysis. In order to perform TEM analysis, we prepared electron-transparent sections of presolar silicate grains by in-situ focused-ion-beam (FIB) lift-out (Stroud 2003, Zega et al. 2007) with an FEI Nova 600 FIB equipped with an Ascend micromanipulator. The small size of the presolar silicates and frequent lack of contrast compared to adjacent matrix material in SE images made the FIB lift-out particularly challenging. Prior to FIB milling, we first overlapped the SE images obtained in the FIB with those acquired in the Auger instrument to precisely locate the correct grain, and then deposited a Pt pillar directly over the grain, followed by a protective carbon mask. The Pt pillar contrasts strongly with the carbon mask in SE images once the milling position reaches the grain, and it also allows the grain to be located rapidly during TEM imaging. With this method we have confidence that if there is basic agreement between the grain dimensions and chemical compositions obtained by Auger and TEM measurements, then the material analzyed in the FIB section is the correct grain, even without additional confirmation by subsequent SIMS measurements on the section.

Transmission electron microscope characterization of FIB lift-out sections of three presolar silicates (grains AH-65a, 166a and 139a) was carried out on a JEOL 2200FS field emission transmission / scanning transmission electron microscope (FETEM / STEM) equipped with a Noran System Six energy dispersive X-ray spectrometer. The spatial resolution of this instrument for imaging is 0.19 nm point-to-point in bright-field TEM mode, and 0.136 nm in high-angle annular dark-field (HAADF) STEM mode. Elemental maps were extracted from EDX spectrum images acquired in STEM mode with nominal 1-nm probe size and spatial drift correction at 30 second intervals. The EDX spectrum from each point in the spectrum images samples the full depth of the FIB lift-out section, ~ 100 nm. To obtain simultaneous high-spatial resolution and high counting statistics on grain 139a, we also performed point-dwell measurements, for which the 1-nm STEM probe was held at fixed positions inside the sample. For quantification of the spectra we used Cliff-Lorimer methods with library k-factors without constraint of the oxygen content to the cation composition. Instrumental k-factors from mineral standards were not used because the grains themselves show significant lateral chemical heterogeneity on a spatial scale smaller than the thickness of the lift-out section, i.e., the uncertainty in chemical composition is dominated by the choice of sampling volume of a heterogenous, non-stoichiometric object, rather than the k-factor accuracy.

## 3. RESULTS

### 3.1. Isotopic Analysis

This study has greatly expanded the number of presolar silicates identified to date (Table 1). The criterion we use to distinguish presolar grains has been described previously (Nguyen et al. 2007b; Zinner et al. 2003). In short, the isotopic ratio of a presolar grain candidate has to be at least $2\sigma_{int}$ outside the $3\sigma$ distribution of isotopic compositions of comparably sized "normal" grains in the same image, where $\sigma_{int}$ is the internal measurement error for the grain (based on counting statistics) and $\sigma$ is the standard deviation of the main distribution. We identified 115 grains in ALHA 77307 and 39 grains in QUE 99177 having O isotopic compositions that satisfy this criterion. According to the $^{28}Si^{-}/^{16}O^{-}$ ratios from the NanoSIMS analysis, 17 of these grains in ALHA 77307 and 5 in QUE 99177 are likely presolar oxides, and the remainders are silicates. Most of the silicates have $^{24}Mg^{16}O^{-}/^{16}O^{-}$ ratios comparable to the surrounding matrix material. Chemical compositions determined by Auger microscopy will be discussed later.

Isotopically anomalous grains can usually be clearly identified in the NanoSIMS isotopic ratio images. Figure 1 shows O isotopic ion images and the $\delta^{17}O/^{16}O$ ratio image for an analysis area from ALHA 77307 containing three < 300 nm grains having O isotopic compositions distinct from the surrounding matrix material. All of these grains are enriched in $^{17}O$ and have normal $^{18}O/^{16}O$ ratios. The correlated $^{28}Si^{-}$ and $^{27}Al^{16}O^{-}$ ion images indicate that grain AH-111a is an Al-rich oxide, and that the other two grains are silicates. The O compositions of all isotopically anomalous oxide and silicate grains identified in this study, shown in Figure 2 and given in Tables 1 and 2, are similar to those seen in previous studies of presolar oxides and silicates (Nittler et al. 2008, and references therein), albeit diluted to a variable and unknown degree due to the aforementioned dilution effects. Of course, this dilution is a concern for all presolar silicate analyses, which all employ ion imaging of densely packed samples, and also for measurement of single grains if they are in proximity to isotopically normal grains.

Grain sizes are estimated from the isotopic images and range from ~180 nm to ~620 nm. For non-round grains, the given diameter is the equivalent diameter of a circle with the same cross-sectional area. The average diameters of presolar silicates and oxides in ALHA 77307 are ~300 nm and ~275 nm, respectively. For QUE 99177, the average diameters are somewhat larger, 330 nm for both silicates and oxides. Note that the finite primary ion beam size typically increases the apparent diameter of most of the anomalous grains in the ion images. As such, the outermost pixels are excluded when manually defining anomalous grains to keep the size inflation minimal. Of course, the grain sizes based on ion images are not always larger than the actual size, and we find that they are in reasonable agreement with diameters estimated from Auger SE images. For this reason we do not attempt to correct diameters estimated from ion images. Moreover, the effective spatial resolution of the images cannot be determined with sufficient precision and the size of a given grain depends somewhat on which pixels are chosen to include in the definition of a grain.

The measured Si isotopic compositions of the presolar silicates are also given in Tables 1 and 2. These ratios are represented in per mil (‰) using the delta notation where $\delta^{i}Si/^{28}Si = [(^{i}Si/^{28}Si)/(^{i}Si/^{28}Si)_{\odot} - 1] \times 1000$. Due to the small grain sizes and lower ionization efficiency of Si in these grains compared to O, the errors are relatively large. We re-measured four silicate grains in ALHA 77307 and three in QUE 99177 for Si isotopes to improve statistics. In Figure 3 we plot the Si isotopic compositions of grains having $\delta^{30}Si/^{28}Si$ errors less than 5%, as well as three other grains anomalous in Si (see below). Also shown is the mainstream correlation line that describes the Si isotopic ratios of presolar mainstream SiC grains, which originate in ~2$M_{\odot}$ AGB stars. Within error the compositions are normal and generally fall above the mainstream correlation line. However, these ratios are also impacted by isotopic dilution.

Analysis of the Si isotopic images for ALHA 77307 also revealed two grains, AH-117b and AH-38, with diameters ~300 nm and ~175 nm that have normal O isotopic compositions, but are depleted in $^{29}Si$ and $^{30}Si$ (AH-117b: $\delta^{29}Si$ = -358 ± 50 ‰, $\delta^{30}Si$ = -202 ± 72 ‰; AH-38: $\delta^{29}Si$ = -281 ± 54 ‰, $\delta^{30}Si$ = -387 ± 61 ‰). These compositions resemble those of presolar SiC grains of type X (Amari et al. 1992) and presolar $Si_3N_4$ grains (Nittler et al. 1995), believed to originate in Type II supernovae. These SiC X grains also tend to have excesses in $^{15}N$ and anomalous C isotopic ratios, but subsequent C and N isotopic analyses were not possible for the two grains from ALHA 77307. The NanoSIMS $^{28}Si^{-}/^{16}O^{-}$ ratios of both $^{28}Si$-rich grains are similar to those of grains that were confirmed by Auger analysis to be silicates, while these ratios are typically much larger for SiC and $Si_3N_4$ grains. Unfortunately, we were unable to locate grain AH-38 in the Auger microscope, and grain AH-117b is yet to be analyzed. The $^{28}Si^{-}/^{16}O^{-}$ ratios and Si isotopic compositions suggest these grains are SN silicates. SiC X grains are quite rare and only make up ~1% of all presolar SiC. If the grains identified in this study are indeed SN silicates, they also make up ~1% of all presolar silicates. Of course, phase determinations based on $^{28}Si^{-}/^{16}O^{-}$ ratios are speculative at best because these ratios are affected by mixing with neighboring material, especially for small grains. The likely phase of these two grains, in light of their O and Si isotopic compositions, is discussed further in section 4.2.2. Grain AH-99 has Si isotopic ratios ($\delta^{29}Si$ = -68 ± 22 ‰, $\delta^{30}Si$ = 113 ± 34 ‰), albeit with relatively large errors, similar to SiC Z grains. SiC grains of type Z also make up ~1% of all presolar SiC and likely condensed in low-mass AGB stars of ~one-third solar metallicity (Hoppe et al. 1997). Since the $^{28}Si^{-}/^{16}O^{-}$ ratio of AH-99 is over 15 times greater than that of the surrounding matrix material, it is most likely SiC.

Thirteen SiC grains (260 to 400 nm in diameter) having anomalous $^{12}C/^{13}C$ and/or $^{30}Si/^{28}Si$ ratios were also identified in the QUE 99177 isotopic imaging survey. Five of these grains were subsequently analyzed for their $^{29}Si/^{28}Si$ ratios (Table 2). One of the SiC grains is a type B grain with a very large enrichment in $^{13}C$ ($\delta^{13}C$ = 15000 ± 3000 ‰), one is a Z grain, and four are mainstream grains. While mainstream (and type B) SiC originate from AGB stars of approximately solar metallicity, type Z grains probably derive from one-third solar metallicity stars. The remaining seven grains are likely to be mainstream SiC as well, though without $^{29}Si/^{28}Si$ data we cannot exclude the possibility that they are type Z. Also identified was a grain having a 30 % depletion in $^{30}Si$ and normal C isotopic composition. The NanoSIMS $^{28}Si^{-}/^{12}C^{-}$ ratio of this grain is greater than that for SiC grains, making it likely that this grain is $Si_3N_4$.

## 3.2. Elemental Distributions: Auger Spectroscopy

A total of 78 presolar grains from ALHA 77307 were analyzed by scanning Auger spectroscopy, including four which were previously identified by Nguyen et al. (2007b). Individual grain analyses were performed for 76 grains, and elemental maps (Fig. 4) were obtained for 9 of these grains and their surrounding matrix. Maps were also acquired for regions surrounding 5 grains that were indistinct in SE images and thus not measured individually. From these maps it is often possible to discern silicates from oxides. The major element concentrations determined for grains analyzed individually are given in Table 3. The spectra for two grains indicate only C, which is likely due to contamination or the C coating on the sample. Several other grains also showed significant C-rich contamination on the surface and the quantification of these grains is less certain than of those free of contamination. The effect of this contamination on the quantification cannot be known precisely, but a significant C signal makes other spectral peaks smaller and the reduced signal-to-noise ratio increases both the uncertainty of abundance determinations and the elemental detection limits. The measurements that are affected by this contamination are noted in Table 3; for these grains the compositional data should only be considered rough estimates.

Four presolar grains were found to be oxides, interestingly all of different type. AH-144 is an Fe-rich oxide, though we cannot determine the stoichiometry because the spectrum for this grain suffers from the C contamination. Though the quantitative results indicate some Fe, grain AH-147b is likely $Al_2O_3$ because elemental maps show this grain is intimately surrounded by very Fe-rich matrix, to which the Fe in the spectrum can be attributed. Grain AH-29 (classified as spinel by Nguyen et al. (2007b) contains Fe and Al with minor Mg and Ni. Grain AH-6 also contains a lot of Al, and minor Mg and Fe. The remaining 68 grains are silicates. Of the 10 grains that were initially classified as presolar oxides according to NanoSIMS analysis and successfully analyzed by Auger microscopy, 6 were found to actually be silicates. These grains are noted in Table 1. We find that ~58% of the silicates have comparable concentrations of Mg and Fe (Mg numbers, or Mg/(Mg+Fe)×100, between 40 and 60), 20% are Mg-rich, and 22% are Fe-rich. Calcium is present in 18% of the silicates, and 9% contain Al. Note that the detection limit for most elements is several atom-%. No correlation exists between Fe/Mg content and the presence of Ca or Al. Nor is there any obvious correlation between elemental composition and isotopic composition.

For several presolar grains, the Auger element maps revealed some fine structure, rims, or heterogeneity in elemental composition. Spot analysis and elemental maps of the 150 nm grain AH-33a reveal Si and O as major elements having the stoichiometry of silica ($SiO_2$), as well as Mg and Fe. From the elemental maps this grain appears to have a 60 nm Mg- and Fe-rich rim, but because the grain is too small to extract for TEM analysis we cannot verify that this "rim" is associated with the grain. Regardless, the minor Mg and Fe content of this grain indicated by the Auger analysis could have come from this surrounding material, rather than being inherent to the grain itself. The ~300 nm silicate grain AH-166a (Fig. 4) has relatively high Ca and Al contents that are slightly displaced from one another, and also appears to be encased in a ~150 nm thick Mg-rich rim. Quantitative analysis indicates this grain is rich in Mg and contains some Fe. The Ca distribution of AH-65a does not appear smooth (Fig. 5) and this grain could have a complex structure. This grain is Mg-rich with no detectable Fe. The latter two grains, which could be structurally interesting, along with the relatively large Mg-rich silicate grain AH-139a were selected for further analysis by TEM.

### 3.3. Mineralogy: TEM

Cross-sections of three presolar silicate grains, all having compositions falling into the Group 1 class, were successfully extracted by FIB lift-out. The need for the Pt pillar in marking the location of the presolar grain is clearly illustrated for grain AH-139a in Figure 6. The boundary between AH-139a and surrounding matrix material, which is also very fine-grained and microstructurally similar to the presolar silicate, is very difficult to discern in either bright-field TEM imaging (Fig. 6a) or HAADF imaging (Fig. 6b). However, the Pt marker allowed us to rapidly locate the grain so that we could confirm its identity through comparison of STEM-EDX maps (Fig. 6) with Auger chemical analysis. The composite RGB elemental map of Mg, Si and Fe reveals a ~500 nm long, and 100 nm thick silicate grain that is relatively rich in Mg and poor in Fe. Point-dwell spectra from the left and right portions of the grain (Fig. 7) show that the grain also contains minor amounts of Al and Ni, and that the chemical compositions of the two sides are distinct, with the left side showing greater Al, Si, Fe and Ni, but less O than the right side. These variations are also apparent in the STEM-EDX maps of O, Al and Ni. Sulfur is difficult to discern, as there is overlap in the peak position from the Mo support used. However, the Mo-L edge contribution to this peak, calculated from the intensity of the higher energy Mo-K edge, allows us to set an upper limit for the S content at 1 at.%. Quantitative compositions obtained from these spectra are shown in Table 4.

Grain AH-166a (Fig. 8) was targeted for TEM analysis because of the apparent compound nature of the grain as seen from Auger elemental maps, with high Ca and Al contents in the interior, greater Si and Mg on the exterior, and a 150 nm Mg-rich rim. The STEM-EDX maps of this grain confirm that the segregation of the Al and Ca content observed in Auger elemental maps of the grain surface (Fig. 4) continues in the cross-sectional direction. In addition, the STEM-EDX mapping reveals that the Al, Ca and Si contents vary greatly at a ~ 30 nm scale, whereas Mg varies more gradually across the grain. These maps further demonstrate that the apparent Mg-rich rim identified in the plan view Auger images does not surround the grain in cross-section and can be attributed to an adjacent Mg-rich grain, rather than a contiguous 150 nm rim. The granularity of the segregation of Al, Ca and Si suggests that this grain is composed of multiple 30 nm scale nanocrystals, possibly Ca, Al oxide interior grains surrounded by Mg-silicate grains. However we could not confirm this with diffraction or lattice imaging due to the lack of strong lattice fringes, and the complex fine-grained nature of the surrounding matrix that complicated diffraction analysis.

The Fe-poor silicate, AH-65a, was targeted because of its relatively large size (~600 nm) and apparent surface heterogeneity. However, in cross-section (Fig. 9) it can be seen that what is left of this grain after the NanoSIMS sputtering has an average depth of only ~30 nm below the surface, and that underlying matrix material may be exposed at the surface in some parts of the grain. STEM-EDX indicates that the composition is non-stoichiometric and that there is variation in composition across the grain, though some of the apparent variation may be due to intrusion of adjacent matrix material. The grain appears amorphous, without discernable diffraction or lattice fringes, which is consistent with the non-stoichiometric composition. However, the possibility of amorphization to a depth of 30 nm during NanoSIMS measurements prevents a definitive analysis of the original grain crystallinity.

These results are in qualitative agreement with the results from Auger spectroscopy for major elements (Table 4). Unfortunately the Auger spectrum of grain AH-139a was severely compromised by C contamination and direct comparison with the STEM results cannot be made. In any case, exact quantitative agreement is not expected for heterogeneous grains because the two techniques sample different volumes, and depending on the scale of the heterogeneity of the grain, neither method gives the true whole grain average composition. Auger spectroscopy sees the top few nanometers of the plan view surface of the grain, with a lateral spatial resolution of several tens of nanometers, whereas STEM-EDX provides greater lateral resolution (down to ~ 1 nm) but samples the full thickness of the lift-out cross-section, ~100 nm. Advantages of the STEM-EDX method are the higher sensitivity to minor elements in the 0.1 to 10 at.% range and the single-nm-scale lateral resolution, which provides a good assessment of the scale of the chemical heterogeneity of a given grain, and thus whether the sampled volume accurately approximates the whole grain. However, Auger analysis can be performed on several grains per hour, whereas the FIB lift-out and STEM analysis of a single in situ silicate grain takes hours to days. In addition, comparison of Auger maps to STEM-EDX maps helps confirm that the extracted FIB slice contains the correct grain. These results clearly demonstrate the importance of Auger spectroscopy to obtain basic compositional information on a large number of grains and to identify interesting candidates for subsequent, much more detailed TEM analysis of select grains in electron transparent cross-sections to deduce the presence of rims, and internal and compound grains. Moreover, the TEM analysis reveals the nature of the matrix material, which in the case of this meteorite is very fine-grained and complex. In light of this, the apparent Mg-rich rim around silica grain 33a could simply be due to neighboring Mg-rich grains, as was the case for grain 65a.

### 3.4. Presolar Grain Abundances

The total analyzed areas in ALHA 77307 of 42880 μm$^2$ and in QUE 99177 of 21170 μm$^2$ were determined by setting a threshold on the summed $^{16}O^{-}$ image of each analysis region. Presolar grain abundances were calculated by dividing the total area of presolar grains by the total area analyzed. As noted above, the grain sizes are derived from the manually defined regions of interest. The matrix-normalized presolar silicate abundance is 161 ± 16 parts per million (ppm), and the presolar oxide abundance is 26 ± 6 ppm in ALHA 77307. The reported errors reflect those from counting statistics (number of identified grains) only. Using the available Auger classifications and assuming the "NanoSIMS oxides" not analyzed by Auger are indeed oxide grains, the abundances become 172 ppm and 16 ppm for presolar silicates and oxides, respectively. However, as 6 out of 10 grains originally classified as oxides based on NanoSIMS measurements were found to be silicates according to the Auger analyses, perhaps a better abundance estimate is obtained if we assume that 60% of the grains classified as oxides based on NanoSIMS measurements are actually silicates. The abundances then become 177 ppm and 11 ppm for silicates and oxides. The abundances of presolar silicates and oxides in QUE 99177 are 140 ± 25 ppm and 20 ± 10 ppm, respectively. No grains from this meteorite were analyzed by Auger spectroscopy, but if we assume that 60% of NanoSIMS classified oxides are misidentified, as was found for ALHA 77307, then the abundances of presolar silicates and oxides become 152 ppm and 8 ppm. The two meteorites have approximately the same abundances of presolar silicates and of presolar oxides, within error. A presolar SiC abundance cannot be meaningfully determined for ALHA 77307 because only a small region was measured for C isotopes. For QUE 99177 we determine an uncorrected abundance of 49 ± 14 ppm for presolar SiC.

The criterion we use to designate presolar grains errs on the side of caution and reported abundances are lower limits. Moreover, as previously discussed, many anomalous grains, especially those with less extreme anomalies, will not be identified by ion imaging due to the dilution of isotopic ratios toward solar. A correction for the Washington University NanoSIMS 50 detection efficiency was applied by Nguyen et al. (2007b) to the calculated abundance of presolar grains. This correction was based on the detection efficiency of presolar spinel grains in densely packed spinel grain residues of mean grain diameters 0.15 and 0.45 μm by ion imaging compared to single grain measurements of these same residues (Nguyen, Zinner, & Lewis 2003). The study showed a severely reduced detection of the smallest presolar grains. Similar systematic studies were not performed on the Carnegie NanoSIMS 50L and the detection efficiency certainly varies between different instruments. Moreover, the detection efficiency will also depend on the specific daily tuning conditions and the software used for image analysis. Nevertheless, the correction should provide to first order a better estimate of the true abundances. Applying the detection efficiency correction described above as a function of grain size and again using the Auger identifications where applicable, the abundance of presolar silicates is 570 ± 150 ppm, and that of presolar oxides is 60 ± 25 ppm in ALHA 77307. Applying the same calculation to the QUE 99177 silicate and oxide grains, the abundances are 400 ± 120 ppm and 60 ± 30 ppm, respectively.

# 4. DISCUSSION

## 4.1. Oxygen Isotopes

The parent stellar sources of presolar oxide and silicate grains, and the nucleosynthetic processes occurring within them, can be deduced from the grains' O isotopic compositions. Presolar oxides have been classified into four groups according to their O isotopes (Nittler et al. 1997b, 1994) and this framework has been used as a basis for discussing the silicate data as well (Floss & Stadermann 2009; Floss et al. 2006; Messenger, Keller, & Lauretta 2005; Messenger et al. 2003a; Mostefaoui & Hoppe 2004; Nguyen et al. 2007b; Nguyen & Zinner 2004; Vollmer et al. 2008). While the group definitions are not exact, they do provide a useful guide for understanding the basic characteristics of the stellar sources and processes that generate the resultant isotopic compositions. Figure 10 shows the O isotopic compositions of presolar oxide grains that were identified by single grain measurements, and also of presolar silicate and oxide grains identified by NanoSIMS ion imaging. The approximate isotopic boundaries of the oxide groups and the GCE trend are also illustrated in this figure. The O isotopic compositions of the presolar silicates fall within the previously observed range, but the effect of isotopic dilution can clearly be seen in comparing the compositions determined by single grain analysis and by ion imaging.

Group 1 grains are believed to have condensed in low-mass AGB stars that have undergone deep convection to mix the products of partial H burning into their envelopes (the first and second dredge-ups). This process results in envelope enrichments in $^{17}O$ and slight depletions in $^{18}O$ (Boothroyd & Sackmann 1999). The predicted surface $^{17}O/^{16}O$ ratio following the first and second dredge-ups is strongly dependent on the initial stellar mass (Boothroyd & Sackmann 1999; Boothroyd, Sackmann, & Wasserburg 1994; Dearborn 1992) and increases with mass to a limit of ~0.004 (~10× solar) in stars of about 2.5 $M_\odot$. The $^{17}O/^{16}O$ ratio in stars more massive than this is also larger than the initial value, but the enrichment is not as large. The first and second dredge-ups are predicted to have a smaller effect on the surface $^{18}O/^{16}O$ ratio, decreasing it from its initial value by up to about 20%. The larger spread in $^{18}O/^{16}O$ observed in Group 1 grains has been interpreted as reflecting the initial metallicities (Z) of the parent stars as determined by GCE (Boothroyd et al. 1994). As the Galaxy evolves the abundances of secondary isotopes (those whose nucleosynthesis depends on metallicity) such as $^{17}O$ and $^{18}O$ increase relative to primary ones (those whose synthesis is independent of initial metallicity) like $^{16}O$ (Clayton 1988; Nittler & Dauphas 2006; Prantzos, Aubert, & Audouze 1996). Thus, higher metallicity stars are expected to have higher $^{18}O/^{16}O$ ratios. The precise relationship between initial isotopic ratios and Z is uncertain but can be explored to a certain extent by analysis of systematic isotopic trends in presolar grains (Alexander & Nittler 1999; Nittler 2005; Nittler & Cowsik 1997). In terms of O isotopes, Boothroyd and Sackmann (1999) calculated the surface composition of red giant stars for a wide range of masses and metallicities under the assumption that the initial $^{17,18}O/^{16}O$ ratios are directly proportional to the Fe/H ratios of stars (GCE line in Fig. 10). For a given mass and metallicity, this model predicts a specific O isotopic composition and thus grain compositions can be inverted to infer masses and metallicities (or initial $^{18}O/^{16}O$ ratios) for the parental stars, albeit in a model-dependent way (Nittler & Cowsik 1997). Comparison of the Boothroyd and Sackmann (1999) models with the O-isotope data for the Group 1 grains that make up the vast majority of the present data set indicates that they condensed in stars of about 1.5 $M_{\odot}$ on average. The origins of Group 1 grains that have $^{17}O/^{16}O$ ratios exceeding the predicted maximum from first dredge-up are still uncertain. However, it has recently been postulated that these compositions could arise from mass transfer in binary systems (Marks, Sarna, & Prialnik 1997; Nguyen et al. 2010b; Nittler et al. 2008; Vollmer et al. 2008). In this scenario, the parent star accreted $^{17}O$-rich material from the intermediate-mass AGB star or nova companion. If these are indeed the parent stars of these highly $^{17}O$-rich grains, then one might expect them to also be enhanced in $^{30}Si$ (José et al. 2004). Only one of the new grains has $^{17}O/^{16}O$ greater than 0.004, but one must keep in mind that the reported ratios are diluted from the actual isotopic ratios.

Grains with $^{17}O$ enrichments comparable to those of typical Group 1 grains, but larger $^{18}O$ depletions, defined as Group 2, are believed to have condensed in low-mass AGB stars that underwent an extra mixing process known as cool bottom processing (CBP; Nollett, Busso, & Wasserburg 2003; Wasserburg, Boothroyd, & Sackmann 1995). During this process, the envelope material is circulated to hotter regions and undergoes H burning. This CBP also explains the low $^{12}C/^{13}C$ ratios of low-mass red giant stars (Charbonnel 1994; Denissenkov & Weiss 1996; Wasserburg et al. 1995) and of presolar SiC grains originating from AGB stars (Alexander & Nittler 1999; Zinner et al. 2006). In intermediate mass stars (> ~4 $M_{\odot}$), the convective envelope extends to the H burning shell and hot bottom burning (HBB) rapidly destroys essentially all $^{18}O$ while producing $^{17}O$ (Boothroyd, Sackmann, & Wasserburg 1995). Only one presolar grain, a spinel, with an isotopic composition suggesting HBB has been identified (Lugaro et al. 2007), but this origin seems unlikely in light of new measurements of relevant nuclear reaction rates (Iliadis et al. 2008). One presolar silicate from the present study was found to have a fairly significant $^{18}O$ depletion consistent with CBP. Of course, the compositions of grains that are depleted in $^{17}O$ or $^{18}O$ are more affected by isotopic dilution than grains with excesses in these isotopes. It is thus inevitable that some of the grains classified here as Group 1 are, in fact, more $^{18}O$ depleted than indicated by the NanoSIMS data and were probably significantly affected by CBP. We should note also that even some of the spread in $^{18}O/^{16}O$ ratios of Group 1 oxide grains for which isotopic dilution is not as important, has been argued to derive from CBP (Nittler 2005).

The Group 3 grains are moderately depleted in $^{17}O$ and most also are depleted in $^{18}O$. Most Group 3 oxides have $^{17}O/^{18}O$ ratios greater than solar and thus plot above the GCE line. These grains are well explained as originating in low-metallicity, low-mass AGB stars whose initial $^{17}O/^{16}O$ and $^{18}O/^{16}O$ ratios were lower than solar due to GCE (Nittler et al. 1997a; Nittler & Cowsik 1997). Interestingly, very few silicates have been found with similar compositions that also indicate a source in low-metallicity, low-mass stars. This points to the possibility that refractory Al-rich oxides are more efficiently made than silicates in such stars. In contrast, most Group 3 silicates (Busemann et al. 2009; Nguyen, Busemann, & Nittler 2007a; Nguyen et al. 2007b; Nguyen & Zinner 2004) and some oxides plot below the GCE line (Fig. 10). The origin of these grains is more ambiguous, since an AGB origin would require that the parent stars had strong sub-solar $^{17}O/^{18}O$ ratios and astronomical evidence suggests that this ratio is already anomalously low in the Sun itself (Penzias 1981; Wilson 1999; Wilson & Rood 1994; Young et al. 2009). Nittler et al. (2008) suggested that these grains might be related to the ($^{18}O$-enriched) Group 4 grains, for which they argued a SN origin, as discussed further below. Indeed, the most $^{16}O$-rich oxide grain in Figure 10 almost certainly condensed in a SN outflow (Nittler et al. 1998).

Group 4 grains are enriched in the heavy O isotopes and many fall along the GCE line. In principle, such grains could come from high-metallicity stars whose initial O-isotopic ratios were heavier than solar. However, dredge-up processes would be expected to have increased the surface $^{17}O/^{16}O$ ratios of such stars such that the grains they produced would no longer fall on the GCE line. On the other hand, $^{25}Mg$ depletions in four presolar Group 4 oxide and silicate grains strongly argue against AGB origins (Nguyen et al. 2010b; Nittler et al. 2008), especially since these stars would have unexpectedly high metallicity. Moreover, origins from Type II supernova have been postulated for several grains having large $^{18}O$ enrichments and near solar $^{17}O/^{16}O$ ratios (Bland et al. 2007; Bose et al. 2010a; Choi et al. 1998; Nguyen et al. 2010b; Nittler et al. 2008), in addition to a Mg-rich olivine ($Mg_{1.66}Fe_{0.34}SiO_4$) grain having the largest $^{18}O$ enrichment found to date and a strong (~1/3 solar) $^{17}O$ depletion (Messenger et al. 2005). In fact, Nittler et al. (2008) performed mixing calculations using a 15 $M_\odot$ supernova model (Rauscher et al. 2002) and determined that the isotopic compositions of three Group 4 oxide grains with $^{25}Mg$ anomalies could be reproduced quite well and argued therefore that most or all Group 4 grains probably originated in supernovae. Moreover, these authors pointed out that the $^{17}O$-depleted Group 3 grains that fall below the GCE line lie near an extension of the SN mixing line that explains the Group 4 oxides and thus may well also be supernova grains with a higher proportion of material from inner $^{16}O$-rich SN layers. Based on these arguments, it is probable that the Group 4 (and possibly Group 3) presolar silicates also originated in SNe. This topic will be further explored in the context of Si isotopes below.

## 4.2. Silicon Isotopes

### 4.2.1. Galactic Chemical Evolution

Low-mass AGB stars of approximately solar metallicity are the proposed parent sources for most of the presolar oxide and silicate grains (e.g. Group 1 and 2 grains). In principle, the same stars could also condense carbonaceous dust, like SiC, later on during their evolution after the stellar atmosphere becomes C-rich from multiple thermal pulses followed by third dredge-up of $^{12}C$ from the He-burning shell. The silicon isotopic compositions of mainstream SiC grains, believed to originate in ~2 $M_\odot$ AGB stars, are described by the "mainstream correlation line" in a three-isotope plot (Fig. 3). The correlation line shown in Figure 3 ($\delta^{29}Si/^{28}Si = -20 + 1.37 \times \delta^{30}Si/^{28}Si$) was derived by Zinner et al. (2007) by fitting the Si isotopic ratios of over 4000 mainstream SiC grains having small errors ($\delta^{29}Si$ errors < 15‰ and $\delta^{30}Si$ errors < 25‰). This correlation is believed to mainly reflect the initial isotopic compositions of the parent stars (Alexander 1993) and, as discussed above for O isotopes, these initial compositions directly reflect GCE (Clayton & Timmes 1997a, 1997b; Gallino et al. 1994; Nittler & Dauphas 2006; Timmes & Clayton 1996). Local isotopic heterogeneities in the ISM likely contribute to the Si-isotope spread (Lugaro et al. 1999), but this cannot fully explain the data, especially when Ti isotopes are also considered (Nittler 2005). In addition to the GCE component, the SiC parent star compositions are also believed to be enhanced in $^{29}Si$ and $^{30}Si$ due to n-capture reactions in the He shell and third dredge-up, but these shifts are not substantial in mainstream SiC (Brown & Clayton 1992; Gallino et al. 1990, 1994; Lugaro et al. 1999; Nittler & Alexander 2003; Zinner et al. 2006).

The Si isotopic compositions of the presolar silicates from this study and others (Mostefaoui & Hoppe 2004; Nguyen et al. 2007b; Vollmer et al. 2008) generally fall along the mainstream SiC correlation line, but are slightly shifted to the left (higher $^{29}Si/^{30}Si$). This is not unexpected as the stellar envelope only becomes enriched in the heavy Si isotopes when the star is C-rich (Lugaro et al. 1999; Zinner et al. 2006). As such, the silicate data should, in principle, better represent the original GCE trend. The silicate data cluster around solar with a range of about 150‰. However, the relatively large error bars for most grains and the potentially severe problem of isotope dilution from neighboring solar system material in the meteorite sections make unambiguous interpretation of the data difficult. The similar range of compositions shared by presolar silicates and SiC does however suggest that their parent stars shared a similar range of metallicities. Thus, the Si isotopic compositions of presolar silicate and SiC grains trace the evolution of the Galaxy as well as the evolution of AGB stars from an O-rich to a C-rich atmosphere.

If the Si isotopic compositions of presolar silicate grains indeed principally represent the initial parent stellar compositions determined by GCE, they should form a positive correlation with the metallicity of Group 1 grains estimated from the O isotopic compositions. Moreover, Ti isotopes provide a method to test these ideas and compare the SiC and silicate data directly. The tight correlation between Si and Ti isotopes ($^{29}Si/^{28}Si$ and $^{46}Ti/^{48}Ti$ ratios in particular) in mainstream SiC grains suggests they both strongly reflect GCE (Hoppe et al. 1994; Lugaro et al. 2001; Nittler 2005), and a similar argument can be made for the relationship between $^{46}Ti/^{48}Ti$ and $^{18}O/^{16}O$ ratios in presolar $Al_2O_3$ grains (Alexander & Nittler 1999; Choi et al. 1998; Hoppe et al. 2003; Nittler 2005). Using these trends, one can calculate the expected correlation between $\delta^{29}Si$ and metallicity (inferred from $^{18}O/^{16}O$ ratios) for presolar silicates. No clear correlation was found in a previous study (Nguyen et al. 2007b) where the O and Si isotopic systems were measured separately. With the larger data set of Vollmer et al. (2008) and the present data, we can re-visit this issue.

Figure 11 shows the measured $\delta^{29}Si/^{28}Si$ values for Group 1 presolar silicates plotted against the inferred initial $^{18}O/^{16}O$ ratios of their parent stars, based on interpolation of first and second dredge up models (Boothroyd & Sackmann 1999). As in Nittler et al. (2008), we consider the initial $^{18}O/^{16}O$ ratio after subtraction of a first dredge-up component rather than an absolute metallicity value, since the precise relationship between O isotopes and Z is unknown. Plotted are the five Group 1 presolar silicates of this study with errors less than 30 ‰ and the data from previous studies (Busemann et al. 2009; Mostefaoui & Hoppe 2004; Nguyen et al. 2007b; Vollmer et al. 2008), again only including grains with errors in Si-isotopic composition smaller than 30 ‰. As discussed later, we draw a distinction based on whether the grains' $^{17}O/^{16}O$ ratios are larger or smaller than $9 \times 10^{-4}$. Based on fits to the $\delta^{29}Si$ and $\delta^{46}Ti$ data for mainstream SiC grains (Alexander & Nittler 1999; Hoppe et al. 1994; Huss & Smith 2007) and to the $\delta^{46}Ti$ and inferred initial $^{18}O/^{16}O$ ratios for presolar $Al_2O_3$ (Choi et al. 1998; Hoppe et al. 2003), we predict a GCE relationship of $\delta^{29}Si = -350 + 0.84 \times ^{18}O/^{16}O$, shown in Figure 11. Even with the larger data set, there is still no convincing trend. Taken at face value this would seem to argue against the GCE interpretation of the Si, O and/or Ti isotope data. However, as previously discussed by Nguyen et al. (2007b), there are a number of major difficulties with the interpretation of Figure 11. First, the ubiquitous problem of isotopic dilution will lead to higher $^{18}O/^{16}O$ and lower $^{17}O/^{16}O$ ratios being measured in many Group 1 and 2 grains than their true compositions. Inferred parental metallicities (initial $^{18}O/^{16}O$ ratios) will consequently be overestimated for true

Group 1 grains, and some Group 2 grains (for which CBP has erased any chance of inferring the initial isotopic composition) will be mis-identified as Group 1 grains, contributing extraneous points to the plot. Second, exclusive of isotopic dilution, if CBP effectively destroyed some $^{18}$O in the parent stars of some Group 1 grains (Nittler et al. 2008), then the inferred metallicities would be skewed to lower values. Finally, the Ti isotopic dataset for $Al_2O_3$ grains is still severely limited, making the calculation of the $\delta^{29}$Si versus $^{18}$O/$^{16}$O relation uncertain.

To semi-quantitatively assess the effect of isotopic dilution on the inferred relationship between Si and O isotopes, we performed some simple mixing calculations (Fig. 12). Previous analysis of simulated NanoSIMS O-isotopic images (Nguyen et al. 2007b) indicated that under typical imaging conditions, grains smaller than 400 nm in diameter experience significant dilution, with some 20-80% (dilution factor, *f*) of the measured O atoms coming from surrounding material. We take as starting compositions the distribution of (Groups 1–3) presolar oxide grains (shown as diamonds in Fig. 12a) found by analysis of isolated grains (Choi et al. 1998; Choi, Wasserburg, & Huss 1999; Nittler et al. 2008; Nittler et al. 1997b; Nittler et al. 1994; Zinner et al. 2003). Since isotope dilution is not a significant problem for such measurements, this can be taken as a reasonable approximation of the true isotopic distribution of presolar O-rich grains. We assume that for Group 1 and 3 grains, the initial Si isotopic composition is related to the initial O isotopic composition by the GCE relation given above. For Group 2 grains, whose original metallicity/O-isotopic ratios cannot be inferred, the initial Si-isotopic compositions are randomly chosen from a uniform distribution in the range $\delta^{29}$Si=0 – 200 ‰. For three dilution factors *f* = 20%, 50% and 80%, we calculated the effect of mixing each grain composition with the appropriate amount of material with terrestrial isotopic composition. A further uncertainty of 25 ‰ was added to each Si isotopic value to simulate typical measurement errors. Finally, using the models of Boothroyd and Sackmann (1999) we inferred initial O-isotopic compositions for parent stars of grains having the diluted O-isotopic compositions and plotted these inferred compositions against the diluted $\delta^{29}$Si values for each grain (Fig. 12b).

The results of the mixing calculations on O isotopes are shown in Figure 12a; the effect of dilution with normal material is obvious. Note that with *f*>50% (typical for all grains smaller than 200 nm), all Group 2 grains are moved into the Group 1 field and the diluted isotopic distribution resembles the observed distribution for presolar silicate grains. Also, grains whose diluted compositions are too close to solar to be identified as presolar grains are excluded from the plots. The effect of isotope dilution on the $\delta^{29}$Si and initial $^{18}$O/$^{16}$O relation is shown in Figure 12b. As the amount of dilution increases, the inferred initial $^{18}$O/$^{16}$O ratio increases and the measured $\delta^{29}$Si value approaches zero. Both effects clearly work to move grains to the right of the (true) initial Si-O isotopic correlation line. Actual highly $^{18}$O-depleted Group 2 grains, indicated by the dotted ellipses, further complicate the picture by adding points to the left of the original correlation line. Comparison of this plot with the presolar silicate data in Figure 11 suggests that the disagreement between the grain data and GCE expectations can at least in part be accounted for by the problem of isotopic dilution in *in situ* measurements. Note that after mixing (*f*>50%) with solar system composition, Group 2 grains have $^{17}$O/$^{16}$O ratios $< 9 \times 10^{-4}$ (Fig. 12a). Interestingly, all of the presolar silicates with $^{17}$O/$^{16}$O ratios higher than this value lie on or to the right of the predicted GCE correlation line in Figure 11. On the other hand, many grains with lower $^{17}$O/$^{16}$O ratios plot to the left of the line, where the mixing calculations predict diluted Group 2 grains to lie (Fig. 12b). Thus, these grains could in fact be mis-identified Group 2 grains. Figure 12b also indicates that the effects of dilution do not necessarily erase the expected correlation for $^{17}$O- and $^{18}$O-depleted Group 3 grains. However, such grains seem to be basically absent from the presolar silicate database (except for those possibly associated with Group 4 grains, see below). In any case, these simple models demonstrate that testing GCE concepts for presolar silicates will require more robust analysis of the O- and Si-isotopic compositions, either through analysis of larger grains or of well-separated grains, such that isotopic dilution becomes unimportant.

4.2.2. Group 4 (and 3) grains: Supernova origins

As discussed earlier, suggested stellar sources of $^{18}$O-enriched Group 4 grains include high-metallicity AGB stars and supernovae. In principle, Si isotopic data for the grains should provide further clues to the origin of these grains and tests of formation scenarios. If Group 4 grains originated in high-metallicity AGB stars, isotopically heavy Si compositions would be expected. Yet the Si isotope data for Group 4 silicates do not indicate such heavy compositions. For example, the three Group 4 grains of this study with relatively high-precision Si isotopic data (AH-33b, AH-147c and QUE-39; Fig. 3) have normal or very slightly (4%) $^{30}$Si-depleted isotopic compositions within errors. Similarly, 18 Group 4 silicates have normal Si isotopic ratios within error, and 3 have $^{29}$Si depletions (Bland et al. 2007; Bose et al. 2010a; Floss & Stadermann 2009; Messenger et al. 2005; Mostefaoui & Hoppe 2004; Nguyen et al. 2010b; Vollmer et al. 2008). As with Mg-isotopic compositions of presolar oxides and silicates (Nguyen et al. 2010b; Nittler et al. 2008), these data do not support a high-metallicity origin for Group 4 grains.

Three of the Group 4 silicates from this study (AH-33b, 87b, and 138d) have O isotopic compositions falling within the range observed for cosmic symplectites, which hold remnants of early isotopically heavy nebular water (Sakamoto et al. 2007; Seto et al. 2008). However, these cosmic symplectites have sizes typically larger than the presolar silicates and are composed of intergrown magnetite and pentlandite (i.e. are rich in Fe, O, and S with minor Ni). These characteristics are inconsistent with the three presolar grains from this study, all of which are Mg-rich silicates, and we exclude the possibility that these grains are cosmic symplectites.

The Si isotopic compositions of SN silicates are more difficult to predict because they depend strongly on the specific mixtures of SN zones that contributed material to the final grain compositions. For example, Figure 13 shows the predicted Si isotopic ratios as a function of radial mass coordinate in the 15$M_{\odot}$ supernova model of Rauscher et al. (2002), which has been used as a basis of comparison in many recent presolar grain studies (Hoppe et al. 2009; Marhas et al. 2008; Messenger et al. 2005; Nguyen et al. 2010b; Nittler et al. 2008). The envelope and H-burnt He/N zone retain the initial Si-isotopic composition of the star, assumed to be solar by Rauscher et al. (2002). However, based on the mainstream SiC data, SN sources of presolar grains likely would have formed with a range of initial Si isotopic compositions, roughly up to $\delta^{29}$Si~$\delta^{30}$Si~200 ‰. Moving inward into the SN ejecta, the $^{18}$O-rich He/C zone (which has experienced partial He burning) is slightly enriched in the heavy Si isotopes due to neutron-capture reactions, the $^{16}$O-rich O/C, O/Ne and O/Si zones are highly enriched in $^{29}$Si and/or $^{30}$Si in variable proportions, and the Si/S zone is depleted in the heavy Si isotopes. Note that Hoppe et al. (2009) recently reported a $^{29}$Si-enriched SiC X grain and showed that its isotopic composition could be quantitatively explained if the $^{29}$Si yield in the O/Si and O/Ne zones were increased by a factor of two, which is consistent with the range of experimental uncertainty for the relevant nuclear reaction cross-sections.

Nittler et al. (2008) showed that the SN mixtures that best explain the Group 4 grains are dominated by material from the outer (envelope, He/N and He/C) zones with only a small admixture of $^{16}$O-rich material from inner zones. In such a mixing scenario, Figure 13 suggests that only modest Si-isotope anomalies might be expected for silicate grains condensed from similar mixtures, with the specific composition depending on which inner zones contribute (and on the initial composition of the star). For example, the mixtures that explain the three oxide grains discussed by Nittler et al. (2008) give Si-isotopic ratios within 150 ‰ of solar, with the exception of hibonite grain KH2 for which a $\delta^{30}$Si value of +400 ‰ is predicted. Thus, the lack of large Si anomalies observed in most Group 4 silicate grains is qualitatively consistent with the proposed supernova origin. It is interesting to note that, although many Group 4 silicate grains have depletions in $^{29}$Si and/or $^{30}$Si, most SN zones have ~solar or isotopically heavy Si compositions. These compositions would be even more isotopically heavy if the initial composition were super-solar, as suggested by mainstream SiC grains.

Although in principle, the Si isotopes could provide constraints on the mixing conditions of the parent supernovae, there are a number of complicating factors. The very wide range of Si compositions, but relatively uniform $^{16}$O-rich composition, predicted for the inner zones of even a single supernova leaves the problem highly unconstrained. Almost any Si-isotopic composition could be matched with the right mixture of zones, without strongly changing the O isotopic composition of the mixture. Further complicating the problem are the relatively large uncertainties of the Si-isotopic measurements, the effect of isotope dilution from surrounding material in the samples, the unknown initial compositions of the parent stars, and the large uncertainties in supernova nucleosynthesis calculations themselves. Clearly progress will require isotopic analysis of additional elements in presolar silicates. For example, presolar silicate grains afford the opportunity to analyze the isotopic systems of Ca, Fe and Mg. Yet measurements of these elements have been severely limited mainly due to the small sizes of these grains coupled with the relatively poorer spatial resolution of the NanoSIMS when analyzing such elements, compared to O and Si. Recent application of FIB milling, however, achieved undiluted Mg isotopic analysis of Group 4 silicates to constrain their SN sources (Nguyen et al. 2010b).

If, as suggested by Nittler et al. (2008), some Group 3 grains are also from supernovae, SN mixtures that would explain their compositions would contain a larger proportion of $^{16}$O-rich inner-zone material than is required to explain the Group 4 data. In this case, larger Si-isotopic anomalies would also be expected, but again the predicted compositions would depend on the specific mixture of zones. Si-isotope data for Group 3 silicates are still scant (Busemann et al. 2009; Nguyen et al. 2007b; this study) and, like other presolar silicates, most have compositions indistinguishable from solar. Two silicates, IDP-G4-4 (Busemann et al. 2009) and AH-98b, appear to be enriched in $^{29}$Si and $^{30}$Si, suggestive of contributions from the O/C or O/Ne zones (Fig. 13), though the ratios for AH-98b have very large errors. Unfortunately, the same problems discussed above for Group 4 grains apply to the Group 3 grains as well, and additional data, especially for other elements, are required to satisfactorily address a possible SN origin for these grains.

We have identified two grains with $^{28}$Si enrichments similar to SiC X grains but with NanoSIMS $Si^-/O^-$ ratios similar to those of silicate grains. The O isotopic compositions of these two grains are normal within 7 %. While we currently do not have sufficient analytical information to explicitly deduce the phase of these grains, we can briefly address the hypothetical possibility that these grains are "X-type" silicate grains. The normal O isotopes of such silicate grains would be difficult to reconcile with SN nucleosynthesis models. Since nowhere in SN ejecta is the gas expected to have solar O-isotopic composition (even the envelope is highly enriched in $^{17}$O from mixing of H-burning ashes), a very specific mixing of layers is required to produce such a composition and the likelihood of such a mixture, out of all possible mixtures allowing silicate condensation, precisely matching the solar composition is very small. Furthermore, several presolar SiC grains found by our NanoSIMS mapping of QUE 99177 have $Si^-/O^-$ ratios similar to silicates, despite the higher $Si^-$ ion yield from SiC than from silicates. Though any abundance determination clearly suffers from limited statistics, we estimate an abundance of 2 ± 2 ppm for the $^{28}$Si-rich grains. For comparison, Davidson et al. (2009) determined the matrix-normalized SiC abundance to be 10 ppm in ALHA 77307. The abundance of SiC X grains would thus be ~100 ppb, given that these grains make up ~1 % of SiC. Our abundance estimate, albeit a bit high, is consistent with this SiC X abundance within error. Based on these observations and the difficulty of explaining solar O-isotopic ratios, we believe that the two ALHA 77307 grains with $^{28}$Si excesses are most likely SiC X grains and/or $Si_3N_4$ grains and that the observed $Si^-/O^-$ secondary ion ratio may be due to contributions from neighboring material.

It is interesting to note that whereas a majority of Group 4 oxide grains fall along the GCE line, a much larger fraction of presolar silicates plot below the line, many with essentially solar $^{17}$O/$^{16}$O ratios. Similarly, while most Group 3 oxides plot above the GCE line, the Group 3 silicate grains plot below this line and have more solar-like $^{18}$O/$^{16}$O ratios. If all the $^{18}$O-rich grains are indeed from supernovae, this indicates a difference in mixing conditions needed to result in compositions that condense oxide phases versus silicate ones. Interestingly, low-density presolar graphite grains also exhibit large $^{18}$O enrichments and close-to-solar $^{17}$O/$^{16}$O ratios (Amari et al. 1995; Jadhav et al. 2006; Stadermann et al. 2005a). The majority of these grains are believed to have condensed in Type II SN, with some proposed to have originated from Wolf-Rayet stars. Of course the details of SN mixing for condensation of C- and O-rich dust differ, but some $^{18}$O-rich material from the He/C zone is necessary in either case. Though silicate absorption features are sometimes prominent in Wolf-Rayet (WR) stars (van der Hucht et al. 1996), whether these features are attributed to circumstellar or interstellar dust is still undetermined. Moreover, O-rich Wolf-Rayet stars are very rare. Other stellar sources observed to have $^{18}$O excesses are R Coronae Borealis and hydrogen-deficient carbon stars (Clayton et al. 2007). However, as with WR stars, the issue remains that these sources are C-rich, so it is unclear how silicates might have formed.

## 4.3. Chemical Analysis

Scanning Auger spectroscopy has only recently been applied to the study of presolar grains (Stadermann et al. 2009; Stadermann, Floss, & Lea 2006a; Stadermann et al. 2005b) but quickly proved useful for assessing the chemical compositions of sub-micrometer grains. Previously, the chemical compositions of six presolar grains from ALHA 77307 were qualitatively determined using this technique (Nguyen et al. 2007b). The present and other recently published studies (Floss & Stadermann 2009; Vollmer et al. 2009b) greatly expand the database of chemical compositions of presolar silicate grains. Figure 14 shows the distribution of Mg/Si and Fe/Si ratios in the presolar silicates from the present study. We find that most of the silicate grains have comparable Mg and Fe contents, while 20 % are Mg-rich and 22 % are Fe-rich. Spectral observations of evolved O-rich stars indicate the presence of Mg-rich crystalline silicates and a larger abundance of amorphous silicates (best matched to a laboratory olivine-composition glass) with higher Fe content (Demyk et al. 2000; Molster et al. 2002b). To explain these observations, Tielens et al. (1998) suggested that crystalline Mg-rich silicates form above the glass temperature and react with gaseous Fe below the glass temperature to form Fe-bearing amorphous silicates. The range of Fe content seen in the presolar silicates could be a reflection of its progressive incorporation. Moreover, if these observations are true, then it might imply that the most Mg-rich presolar silicate grains are crystalline, while the large remainder is amorphous with more Fe. On the other hand, models of grain condensation predict the formation of Mg-rich silicates under equilibrium conditions whereas more Fe-rich silicates will condense under non-equilibrium conditions (Ferrarotti & Gail 2001). Both secondary reaction with Fe gas and non-equilibrium condensation are viable explanations for the Fe seen in the silicate stardust. Alternatively, the Fe may not be inherent to the dust forming disk, but may rather be a consequence of parent body or nebular alteration (Nguyen & Zinner 2004). Indeed, presolar silicate grains from the thermally altered meteorite Adelaide have much higher Fe contents (Floss & Stadermann 2010) than those from the less altered CR chondrites (Floss & Stadermann 2009) and ALHA 77307. These highly primitive meteorites have suffered negligible thermal alteration, however, and some of the Fe observed in the presolar silicates is likely innate. This is the case for presolar silicates from Acfer 094 having anomalous Fe isotopic compositions (Mostefaoui & Hoppe 2004; Vollmer & Hoppe 2010).

Although olivine is the most stable crystalline silicate condensate (Ferrarotti & Gail 2001) and is abundantly observed around evolved stars (Demyk et al. 2000), and amorphous solids of olivine composition are also abundant in the interstellar medium (Kemper, Vriend, & Tielens 2004), we find that the majority of the presolar silicates do not have stoichiometric compositions consistent with olivine. In fact, the chemical make-up of the majority of presolar silicates is nonstoichiometric. Of the 55 presolar silicates for which we have reliable quantified Auger results, ten grains have compositions similar to olivine, with one probable forsterite (Mg-rich endmember), and five have compositions similar to pyroxene ($(Mg,Fe)SiO_3$). This observation of abundant non-stoichiometric silicate grains is substantiated by the TEM analyses. In contrast to the number of grains analyzed by Auger spectroscopy, only 27 presolar silicates, including those presented herein, have been analyzed by TEM due to the involved sample preparation techniques and the small sample volume remaining after SIMS analysis. Of the 15 grains identified in meteorites, ten are non-stochiometric and amorphous to weakly nanocrystalline (Nguyen et al 2007b; Vollmer et al. 2009a; this study), three grains contain detectable crystalline olivine (Stroud, Floss, & Stadermann 2009; Vollmer et al. 2009a), one grain is reported as crystalline $MgSiO_3$ with a perovskite structure (Vollmer et al. 2007), and one is amorphous $MgSiO_3$ (Nguyen et al. 2010a).

The mineralogies of the meteoritic presolar silicates appear to differ from those of presolar silicates in IDPs characterized by TEM, which consist of three crystalline olivine grains (Busemann et al. 2009; Messenger et al. 2003a, 2005), three equilibrated aggregates (Keller & Messenger 2008; Messenger et al. 2010), and six glass with embedded metal and sulfide (GEMS) grains (Floss et al. 2006; Keller & Messenger 2008; Messenger et al. 2003a). Identified as a major constituent in IDPs (up to 50 wt. %), GEMS are characterized by non-stoichiometric but approximately chondritic (solar) chemical compositions that vary on a nanometer scale (Bradley 1994). Only a few percent of GEMS grains have anomalous isotopic compositions which verify the grains' circumstellar origins (Keller & Messenger 2008). The amorphous silicate grains that we have studied by TEM also exhibit very fine-scale heterogeneity and non-stoichiometric compositions, but not the distinct tens-of-nanometers-scale metal or sulfide subgrains characteristic of most GEMS. Two presolar silicates from Acfer 094 were reported with small metal subgrains and low S contents (< 1 at.%), but the spatially correlated distribution of the metal subgrains is not characteristic of GEMS and suggests that the metal grains are parent body alteration products (Vollmer et al. 2009a). Also reported was an amorphous presolar silicate of near chondritic composition with multiple sulfide inclusions (Vollmer et al. 2009a). However, the isotopic signature of this grain could not be confirmed and the identification as a presolar meteoritic GEMS remains tentative.

Detailed chemical comparison, in addition to microstructure, can provide some insight into the similarities and differences of amorphous meteoritic silicates to bona fide GEMS grains. Auger analysis indicates that many of the presolar silicate grains have sub-solar Mg/Si and Fe/Si ratios, similar to what is observed for GEMS grains in IDPs (Fig. 14; Keller & Messenger 2004). Moreover, the distribution of Mg/Si (and Fe/Si) in the presolar silicates is similar to that seen in GEMS (Fig. 15), although the lack of S detection by Auger analysis suggests that S is greatly reduced compared to the GEMS average. GEMS grains also have roughly chondritic Al contents, whereas Ca and S are both depleted relative to solar (Keller & Messenger 2004). These concentrations are below the detection limit of the Auger nanoprobe, but are observable by STEM-EDX. Thus, a GEMS classification for silicate grains analyzed by Auger alone cannot be ruled out completely based on the lack of detectable Al, Ca or S. However, a GEMS classification can be definitively ruled out for the thirteen silicate grains (Table 3) that show Ca and/ or Al significantly above the published concentration distributions for GEMs in IDPs (Ca/Si $\leq 0.07$; Keller & Messenger 2004). For the three grains analyzed in this study by TEM, the EDX compositional data (Table 4) show that AH-166a and AH-65a are significantly enriched in Ca (Ca/Si = 0.22 and 0.10, respectively) and Al (Al/Si = 0.47 and 0.21, respectively) relative to solar, but that AH-139a has near-solar Al (Al/Si = 0.095) and sub-solar Ca (Ca/Si = 0.005). The Mg content of AH-139a (Mg/Si = 1.30) is above solar values and the S content is no higher than 1 at.%. These values fall within the range observed for GEMS in IDPs (Keller and Messenger 2004), and thus this grain is a meteoritic presolar GEMS candidate.

Whether or not some meteoritic presolar silicates are GEMS grains, the fact that to date no amorphous presolar silicates with significantly super-solar Al or Ca have been reported in IDPs and that all of the reported meteoritic presolar silicates are S-poor relative to the average GEMS grain suggests that there are real differences between the meteorite and IDP presolar silicate grain populations. How much of the difference is due to statistical fluctuations of small data sets, sampling bias and/or instrumental bias, and how much reflects cosmochemical processes is yet to be determined. The TEM data to date for meteoritic grains are biased by the targeted selection of larger and more "interesting" grains, e.g., those with possible rims. The low average S content of the meteoritic grains determined from Auger could be due in part to preferential sputtering of sulfides during SIMS measurements. Additional TEM analysis of more representative grains could address this point by revealing the presence or absence of sulfide inclusions in the cross-sectional direction, at depths not affected by SIMS. Selection bias or statistical fluctuations may also affect the IDP GEMS data. For example, most of the GEMS with S contents in the 1 at.% range interestingly came from one IDP (Keller & Messenger 2004).

Despite any biases in the available data, cosmochemical processes play the dominant role in determining the grain microstructures and compositions, and likely produce some of the observed differences between the IDP and meteoritic grain populations. GEMS grains display a range of S content, and some are S-poor. The sulfidization of Fe in GEMS grains in the solar nebula has been invoked to explain this compositional range (Keller & Messenger 2004). If nebular alteration indeed produced S-rich GEMS, then this would imply that meteoritic presolar silicates escaped similar sulfidization. It is possible that asteroids and comets sampled different collections of presolar silicates, and that more extensive nebular and/or parent body processing in meteorites has altered the chemical nature of some of the grains. As described in section 4.4, the overall abundance of presolar silicate grains is lower for meteorites that display evidence of hydrothermal alteration, which indicates grains are destroyed by parent body processing. However, the grains identified as presolar are those that survived any parent body processing without gross dilution of a circumstellar isotopic signature, and presumably experienced limited chemical exchange of at least Si and O. Parent body alteration could account for the higher Fe contents of some meteoritic presolar silicates, as indicated by the two metal subgrain-bearing silicates reported by Vollmer et al. (2009a), and annealing of amorphous species could result in weakly crystalline grains, though this might also result in isotopic exchange of O. It is unlikely, however, that the meteoritic silicates are the result of aqueous or mild thermal alteration of the presolar silicate species identified in IDPs. Alteration experiments performed on anhydrous IDPs show that GEMS grains become poorly crystalline but retain their chemical compositions and enstatite grains become phyllosilicates (Nakamura-Messenger et al. 2007). In fact, amorphous silicate matrix material in the primitive carbonaceous chondrite Acfer 094, which is also abundant in silicate stardust and has evidence of hydration, contains inclusions of Fe-sulfides and metal grains and likely derived from processed GEMS grains (Keller, Nakamura-Messenger, & Messenger 2009). Thermal alteration of GEMS above 700°C has been found to result in transformation to Fe-rich silicates with loss of fine grained metal (Brownlee et al. 2005), but such high temperatures were not experienced on the meteorite parent bodies and would certainly affect the grains' isotopic signatures and petrographic relationship to adjacent fine-grained matrix material. Thus, if the precursors of meteoritic presolar silicates were indeed silicates from IDPs, one would expect to find evidence of GEMS-like inclusions or identify a phyllosilicate structure.

There is of course the possibility that the observed chemical compositions of some presolar silicate grains are not attributable to secondary alteration, but rather are primary features. This is evidenced by the presence of Mg-rich crystalline grains and more Fe-rich amorphous grains, in agreement with inferences from astronomical observations. The Mg-rich presolar silicate grains with olivine crystal structures are consistent with formation by equilibrium condensation, similar to the majority of presolar oxide grains, which are single crystals absent of subgrains: $Al_2O_3$ (Stroud, Nittler, & Alexander 2004), hibonite (Stroud, Nittler, & Alexander 2008; Stroud et al. 2005, Zega et al. 2006), and spinel (Zega et al. 2010). The available Auger and TEM data do not support the claim that internal subgrains of $Al_2O_3$ or $TiO_2$ seed nuclei are required for the formation of silicates (Demyk et al. 2000; Ferrarotti & Gail 2001) or oxides. Moreover, these findings do not corroborate the suggestion that accretion of Fe metal onto silicate grains during condensation would result in silicates with platy Fe inclusions (Kemper et al. 2002). One peculiarity is the scarcity in the presolar grain database of crystalline pyroxene, predicted to be the most abundant crystalline silicate species in evolved stars (Molster et al. 2002b). Of course the struggle lies in spectral interpretation and the low statistics of TEM studies.

If the chemical compositions of the grains were inherited during initial condensation, it is clear that the majority of the silicates, excluding the Mg-rich olivines, were not produced by direct equilibrium condensation, which neither explains the observed range of compositions nor the amorphous structures. Amorphous or weakly nanocrystalline grains with non-stoichiometric compositions can condense directly by rapid undercooling of the circumstellar gas envelope to supersaturation, i.e., non-equilibrium conditions. However, it is difficult to explain the internal chemical heterogeneity of the grains by direct condensation from a gas of fixed or slowly varying composition. Subsequent annealing, possibly during the collapse of the molecular cloud to form the solar nebula, and/or radiation processing in the ISM could produce chemical heterogeneity as a result of internal diffusion and solid-state precipitation of compositionally distinct nanoscale-subgrains (Sun et al. 2004). It is also possible that non-stoichiometric grains could be produced by a multistep condensation process, in which oxides are coated with granular silicate rims, or grains condensed at different times or temperatures aggregate in the circumstellar envelope before reaching the ISM. Grain AH-166a (Figs. 4, 8), which shows Al and Ca segregation to the grain center, and more Mg and Si at the edge, could have formed by such a process.

The discovery of a presolar silica grain is unusual and only three others have recently been identified (Floss & Stadermann 2009; Bose et al. 2010b). Silica has tentatively been observed in the O-rich dust shells around evolved stars (Molster, Waters, & Tielens 2002a). There are several circumstances under which $SiO_2$ is predicted to condense. It is the major condensate in very low metallicity stars that do not have much Mg at all, but the O isotopic composition of AH-33a does not indicate such a low Z source. Calculations of condensation under both equilibrium and non-equilibrium conditions indicate that $SiO_2$ (as well as enstatite and Fe) is a significant condensate in stellar atmospheres with Mg/Si < 1 (Ferrarotti & Gail 2001). Some F and G stars have been observed to have Mg/Si ratios that fit this criterion, but they are rare (Reddy, Lambert, & Allende Prieto 2006; Reddy et al. 2003). Recent laboratory annealing experiments have also demonstrated that cristobalite, a silica polymorph, can form by heating of a glass with enstatite composition to 990°C (Roskosz et al. 2009).

### 4.4. Abundance Comparisons

Presolar silicate grains have been abundantly identified in IDPs (Floss et al. 2006; Messenger et al. 2005; Messenger et al. 2003a) and various primitive meteorites (Floss & Stadermann 2009; Mostefaoui & Hoppe 2004; Nagashima et al. 2004; Nguyen, Alexander, & Nittler 2008; Nguyen et al. 2007b; Nguyen & Zinner 2004; Vollmer et al. 2009b). In each case the least altered samples were chosen for analysis because silicate grains are destroyed by aqueous alteration. Thus, most of the meteorites show exceptionally primitive characteristics and are not representative of their meteorite class. Figure 16 illustrates the abundances of presolar phases in the meteorites ALHA 77307, MET 00426, and QUE 99177 and in IDPs. These studies were conducted by ion imaging in the NanoSIMS and the reported abundances are uncorrected for the detection efficiency.

Floss et al. (2006) studied a suite of IDPs and determined an average presolar silicate abundance of ~120 ppm for all samples, and an abundance of ~375 ppm in a supposed isotopically primitive class. This class of IDPs had bulk N isotopic anomalies and was considered "isotopically primitive" due to the presence of $^{15}$N-rich hotspots, some C isotopic anomalies, and abundant presolar silicates. Recently, several IDP samples collected in the dust stream of comet Grigg-Skjellerup were found to be exceptionally primitive with presolar silicate abundances up to 1.5 % (Busemann et al. 2009; Nguyen et al. 2007a). These abundances far exceed those in meteorites. The meteorites containing the highest concentrations of presolar silicates thus far are Acfer 094 (unclassified) (163 ppm; Vollmer et al. 2009b), MET 00426 (CR) (120 ppm; Floss & Stadermann 2009), QUE 99177 (220 ppm; Floss & Stadermann 2009) and ALHA 77307 (177 ppm; this study). The clear consequence of parent body alteration on presolar silicate survival is demonstrated by the comparatively low abundance in the slightly altered ordinary chondrites Semarkona and Bishunpur (15 ppm; Mostefaoui & Hoppe 2004; Mostefaoui et al. 2003). These abundances indicate that IDPs are generally more primitive than any meteorite analyzed to date. However, the H and N isotopic compositions of CR chondrites have been likened to those of IDPs (Busemann et al. 2006; Messenger et al. 2003b), and the high presolar silicate abundance in QUE 99177 substantiates this relation. The abundances of presolar O-rich phases in ALHA 77307 are comparable to those in the aforementioned CR chondrites, but the matrices of these CR chondrites have more anomalous H and N isotopic compositions than those of ALHA 77307 (Alexander et al. 2007).

Because oxide phases are not as susceptible to parent-body alteration as are silicate grains, one might expect the abundance of presolar oxides to be similar among these primitive meteorites and IDPs. However, hitherto the database contests this belief. The presolar oxide abundances in Acfer 094 and ALHA 77307 are 26 and 11 ppm, respectively. Upper limits of 4-5 ppm were determined for the CR chondrites MET 00426 and QUE 99177 by Floss and Stadermann (2008), while we obtain an abundance of 8 ppm for the latter meteorite. Interestingly, only one presolar aluminum oxide grain has been identified in IDPs (Stadermann, Floss, & Wopenka 2006b), though the amount of IDP material analyzed is comparatively smaller than meteorites. The estimated abundance of 600 ppm is highly suspect because it is based on one data point. The lower abundance of presolar oxides in MET 00426, QUE 99177, and perhaps IDPs is unexpected and implies that either the parent bodies of ALHA 77307 and Acfer 094 received a greater proportion of presolar oxides than the CR chondrites and IDPs, or that the destruction of presolar oxides was much larger in the latter samples. However, apparently neither the parent bodies of CR chondrites, nor those of anhydrous IDPs experienced any significant thermal processing. This conundrum may be clarified by examining the SiC concentrations.

The abundance of presolar SiC grains had previously been established in various meteorites based on noble gas analyses to range up to ~30 ppm (Huss & Lewis 1995; Huss et al. 2003), and SiC X and Z grains each make up about 1% of all SiC. As only three SiC were identified in this study of ALHA 77307 and many presolar SiC grains were surely missed without analysis of C isotopes, any abundance determination is inconsequential. However, Huss et al. (2003; noble gas analysis) and Davidson et al. (2009; NanoSIMS analysis) report a SiC abundance of ~9-10 ppm in ALHA 77307. Only one SiC grain has been identified in IDPs (Stadermann et al. 2006b). Huss et al. (2003) attributed the variation of presolar grain abundances to thermal processing in the solar nebula. If this were the case, one would expect meteorites of the same class to have similar abundances of presolar SiC (and oxides) and for the abundances to scale with the degree of thermal processing (bulk compositions). Yet, while CR chondrites accreted moderately processed components (Huss et al. 2003) both CR chondrites MET 00426 and QUE 99177 have very high SiC concentrations of ~30 – 110 ppm (Floss &

Stadermann 2008) and 55 ppm (this study), respectively. Recent determinations of the SiC abundances in other CRs also report a high range of 27 ppm – 55 ppm, interestingly with Renazzo having an abundance of 36 ppm (Davidson et al. 2009). These data do not argue for alteration of presolar grains due to nebular processing. Moreover, because both oxide and SiC grains are quite resilient, one might observe a correlated variance of presolar grain abundances among the samples studied if there is a true relation between presolar grain survival and degree of nebular processing. Yet while the concentration of presolar oxides in ALHA 77307 is greater than that in the other samples, the abundance of SiC is much lower (Fig. 16). Thus, the abundance variation of presolar oxide and SiC grains in different meteorites does not point to pre-accretionary thermal processing or parent body alteration effects. On the other hand, oxidation in the hot ($T \geq 900°C$) solar nebula would destroy SiC grains in less than several thousand years, suggesting that surviving SiC accreted after the inner nebula cooled below 900°C or in the outer regions of the solar nebula (Mendybaev et al. 2002). Partial oxidation is evidenced by the morphologies of some pristine presolar SiC grains (Bernatowicz et al. 2003). This oxidation would not have affected presolar oxides, however, and the distribution of presolar grains in the early solar nebula was likely not uniform. Further systematic studies would help to clarify the issue.

## 5. SUMMARY AND CONCLUSIONS

We have presented O and Si isotopic data for presolar silicate, oxide, and SiC phases identified in the carbonaceous chondrites ALHA 77307 and QUE 99177. The O compositions of the silicate grains span all four groups of presolar oxide grains, with the majority having origins in low-mass red giant and AGB stars, as expected. We find that the isotopic compositions of the rare Group 3 and 4 presolar silicates differ from those of Group 3 and 4 oxides. Many of the Group 4 silicates have smaller $^{17}O/^{18}O$ ratios than Group 4 oxides. Moreover, while the compositions of Group 3 oxides correspond well with low-metallicity sources, those of the Group 3 silicates could not be produced in such stars and it is possible that these silicate grains share the same heritage as the Group 4 silicates (and oxides). Multi-isotopic analysis of two Group 4 oxide grains led to the inference of a SN origin for most Group 4, and 3, grains (Nittler et al. 2008). Indeed, the sub-solar $\delta^{30}Si$ ratios of three Group 4 silicates from this study also argue for SN sources. Origins in Wolf-Rayet stars and R Coronae Borealis, whose extreme $^{18}O$-rich compositions mimic those of some Group 4 silicates, cannot yet be excluded, however these stars are C-rich and condensation of O-rich phases is hard to reconcile. Presuming that the Group 4, and some Group 3, oxides and silicates condensed in SN, the distinctive O isotopic compositions could be attributed to specific mixing conditions in SNe, notwithstanding the chaotic nature of SN mixing. The sensitivity of the condensing grain phase to the chemical composition of the gas may necessitate distinct zone mixtures for silicates and oxides, resulting in their divergent isotopic compositions. Though SN zone mixtures for presolar oxides and silicates thus far appear qualitatively similar (i.e., most material derives from the outer layers with small amounts of inner zone material; Messenger et al. 2005; Nguyen et al. 2010b; Nittler et al. 2008), a real difference in the zone proportions may be present. This is currently difficult to assess because the supernova model fits are not unique and large uncertainties exist with the supernova nucleosynthesis calculations.

Both Auger spectroscopy and TEM analyses conclude that the majority of presolar silicate grains in meteorites are chemically heterogeneous on a fine scale and have non-stoichiometric compositions. Moreover, though Mg-rich compositions are expected from condensation calculations and circumstellar observations, most of the presolar silicates have comparable Mg and Fe contents. While the high Fe content of some grains is likely a primary characteristic, secondary reaction or alteration has also clearly been shown to affect the Fe content. Chemical analysis of the presolar silicates from QUE 99177 (Floss & Stadermann 2009), which is less altered than ALHA 77307, and a few from primitive anhydrous IDPs (Floss et al. 2006) also reveals Fe-rich compositions, suggesting that the Fe-content of some silicates is indigenous. Yet, most presolar silicates identified in IDPs are Mg-rich (Busemann et al. 2009; Floss et al. 2006; Keller & Messenger 2008; Messenger et al. 2005; Messenger et al. 2003a), pointing either to secondary alteration effects on the chemical compositions of silicates in meteorites, or to different collections of silicates sampled by the parent bodies of meteorites and IDPs. Further evidence for the latter are the predominance of presolar crystalline silicates and GEMS in IDPs, and the lack of these species hitherto in meteorites.

The abundances of presolar silicates generally provide a gauge for the degree of nebular and parent body processing. The observed presolar abundance variation among different meteorites of the more resilient oxide and SiC phases, however, does not correlate with degree of processing. This variation suggests that mixing in the solar nebula was not uniform and that there was an inhomogeneous distribution of presolar phases. Unfortunately, statistically meaningful presolar oxide and SiC abundances in IDPs are not currently available to further investigate this point.

*Acknowledgements* – The work of LRN was supported by NASA grants NNG04GF61G and NNX07AJ71G. The work of FJS was possible through NASA grants NNX08AI13G and NNX07AI82G. RMS would like to acknowledge the NASA Cosmochemistry and LARS programs (grants NNH09AL20I and NNH07AG02I), as well as the Office of Naval Research. AN acknowledges support from the Carnegie Institution of Washington. We are grateful to Scott Messenger for a helpful and positive review.

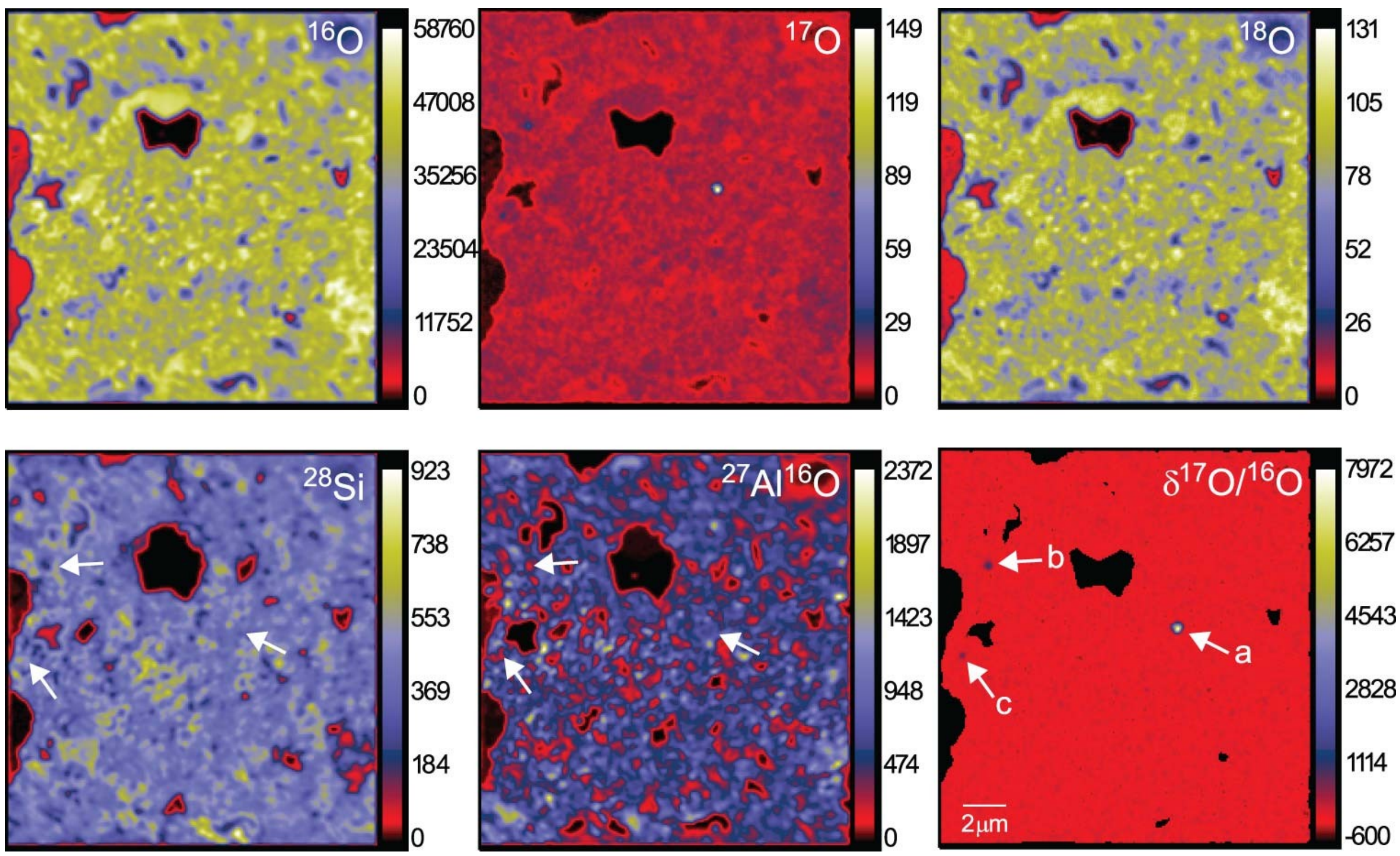

Figure 1. Ion images of the three O isotopes, $^{28}Si$, and $^{27}Al^{16}O$ for a 20 × 20 μm$^2$ region of the ALHA 77307 thin section. Images are scaled to the brightest pixel, and the color bars give ion counts per second. The corresponding ratio image, where the ratios are given as delta values, reveals 3 isotopically anomalous grains (AH-111a, b, and c, designated in the figure by the corresponding letter) that are all enriched in $^{17}O$. None of these grains were analyzed by Auger spectroscopy, but the NanoSIMS measurement indicates grain 111a is an Al-rich oxide, and grains 111b and 111c are silicates.

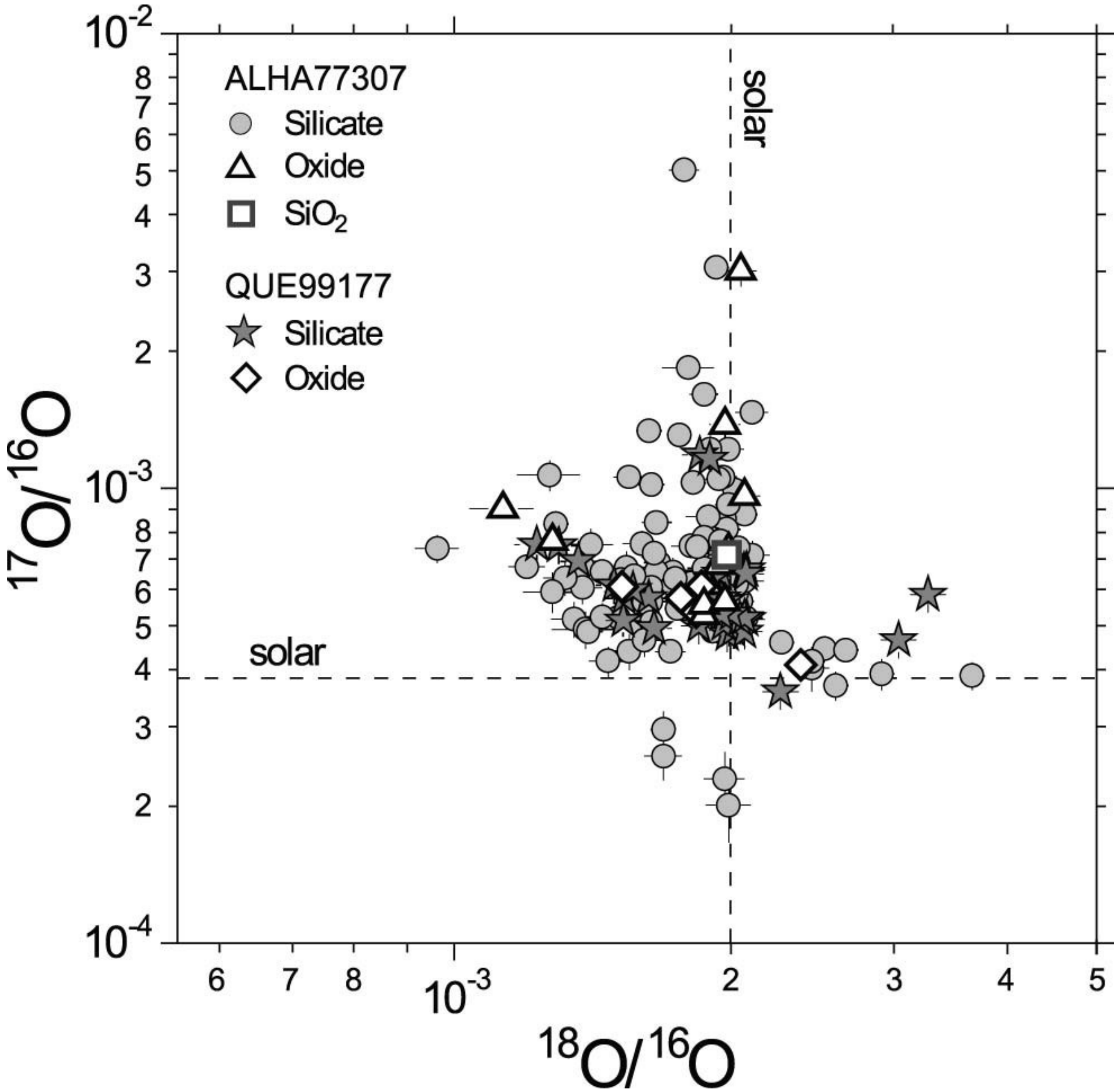

Figure 2. Oxygen isotopic compositions of the presolar silicates, oxides, and $SiO_2$ identified in this study. The majority of the grains have enrichments in $^{17}O$ and depletions in $^{18}O$. Error bars are 1σ. The dotted lines indicate the isotopic ratios of solar system materials ($^{17}O/^{16}O = 3.83 \times 10^{-4}$; $^{18}O/^{16}O = 2.01 \times 10^{-3}$).

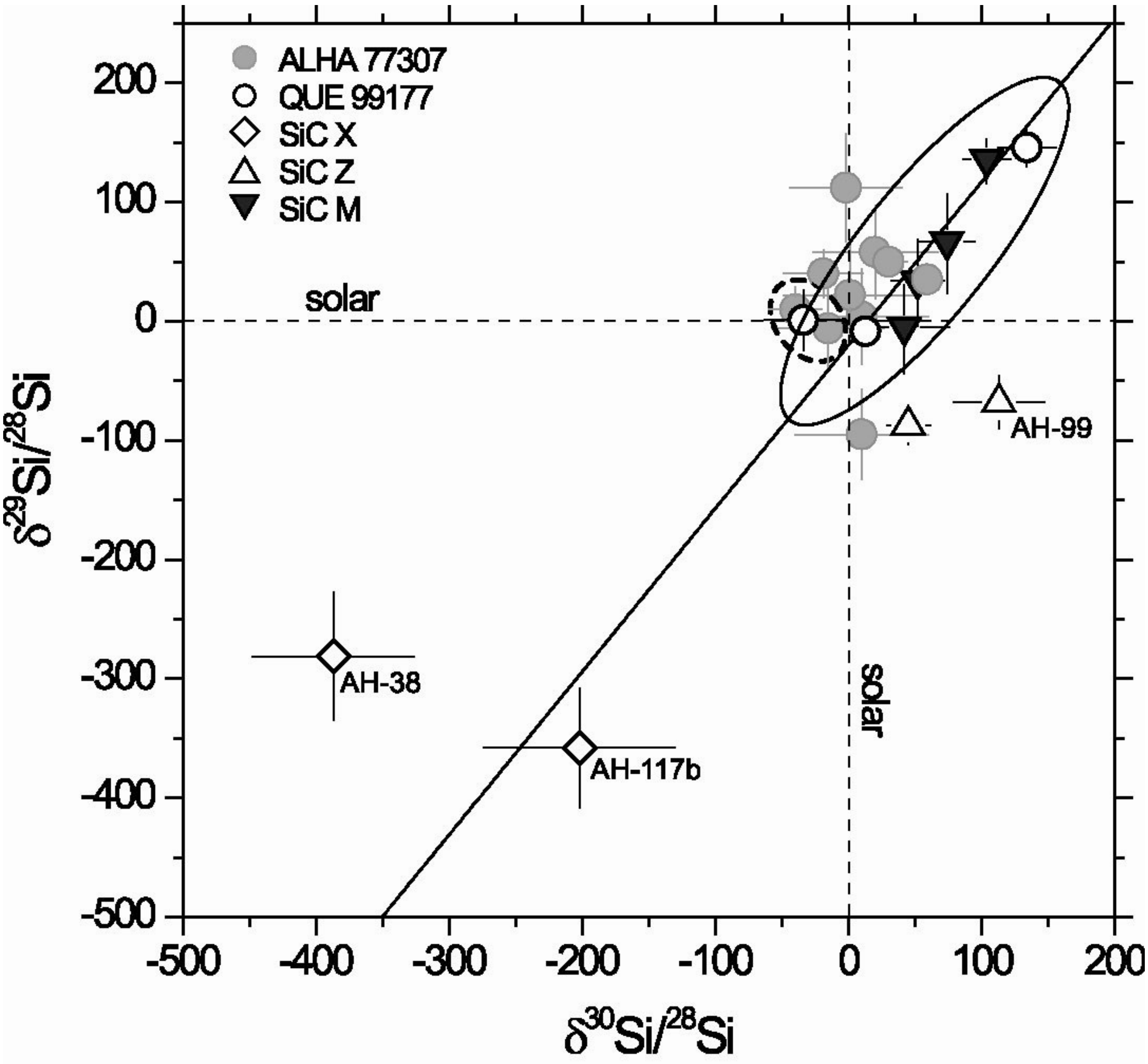

Figure 3. Silicon isotopic compositions of presolar silicate grains having $\delta^{30}$Si errors less than 50 ‰, and of the SiC grains identified in this study. These ratios are expressed as deviations away from solar in permil (‰). The dashed lines indicate solar compositions. The "mainstream correlation line" (solid) depicts the initial parent stellar compositions and the effects of GCE (Zinner et al. 2007). The silicate grains fall within the distribution of mainstream SiC grains (ellipse) but generally are lighter in $^{30}$Si. These grains are classified as Group 1 grains, save for the three Group 4 grains (AH-33b, AH-147c, and QUE-39) enclosed by the dotted ellipse. The isotopically light Si compositions of the Group 4 grains argue against high metallicity stellar sources.

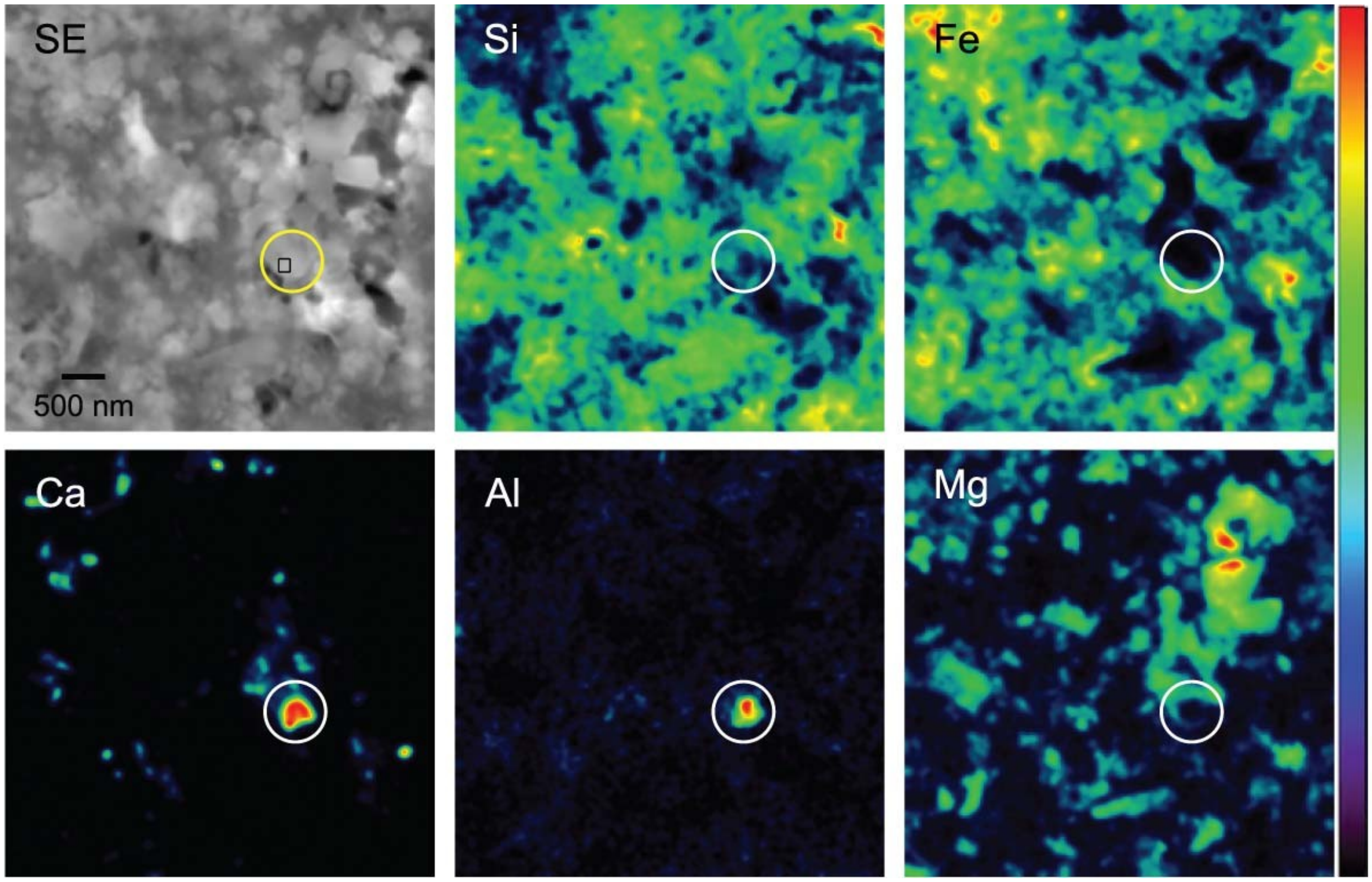

Figure 4. Scanning Auger spectroscopy elemental and secondary electron maps of a 5×5 μm$^2$ area containing presolar silicate AH-166a. The small box in the SE image indicates the portion of the grain that was analyzed quantitatively. This grain contains substantial amounts of Al, Ca, and Mg, and also has some Fe. The thin Mg-rich "rim" encircling the grain was determined by later FIB-TEM analysis (see Fig. 8) to be neighboring grains rather than an inherent feature.

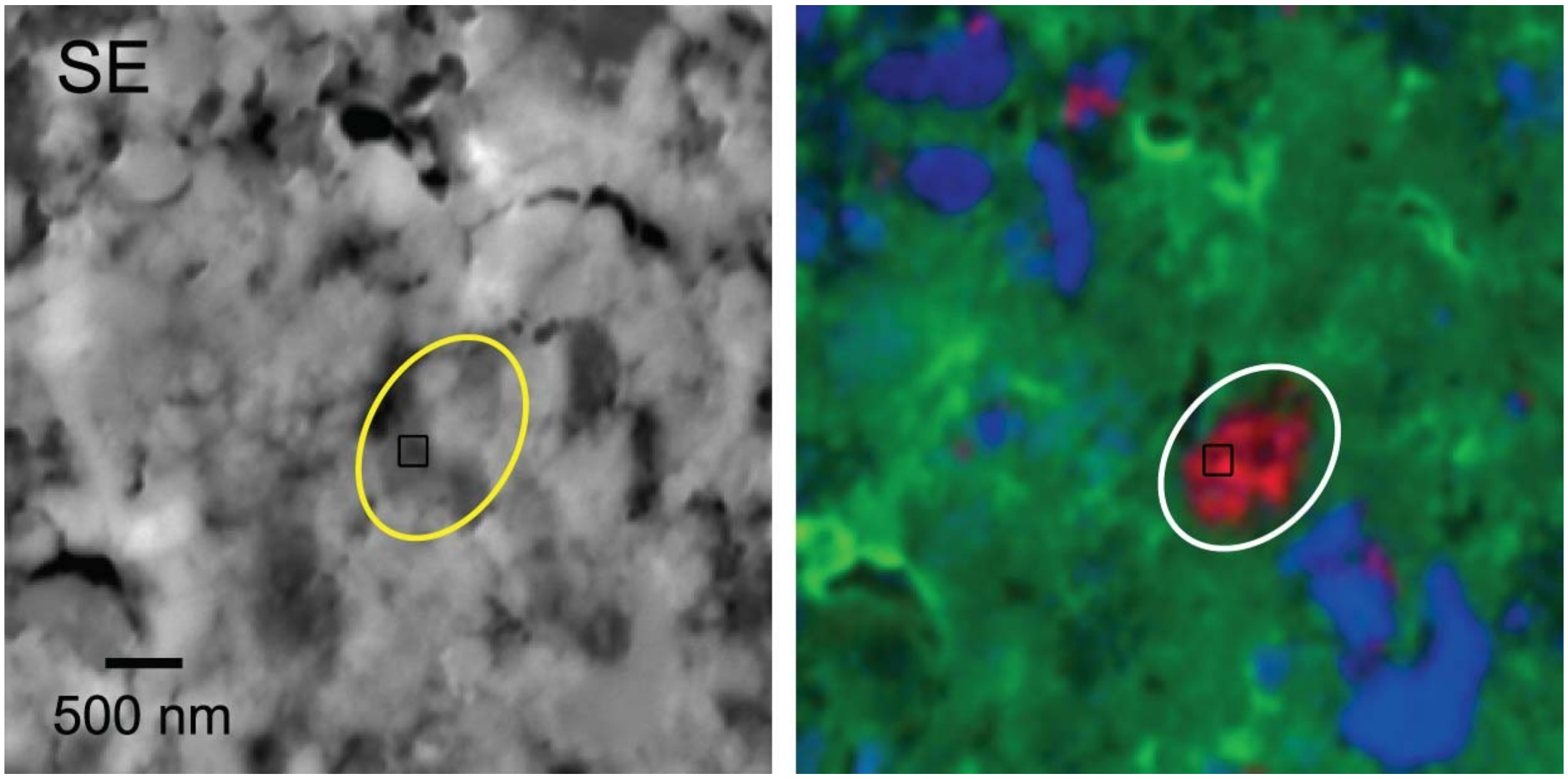

Figure 5. Secondary electron and composite elemental map obtained with the Auger nanoprobe of a 5×5 μm$^2$ region containing grain AH-65a (circled). In the composite map, red is Ca, green is Fe, and blue is Mg. From these maps, the Ca-rich grain looks to have some surface topography and non-uniform Ca concentration. The small box indicates the portion of the grain that was analyzed quantitatively. This grain was extracted and analyzed by FIB-TEM (Fig. 9).

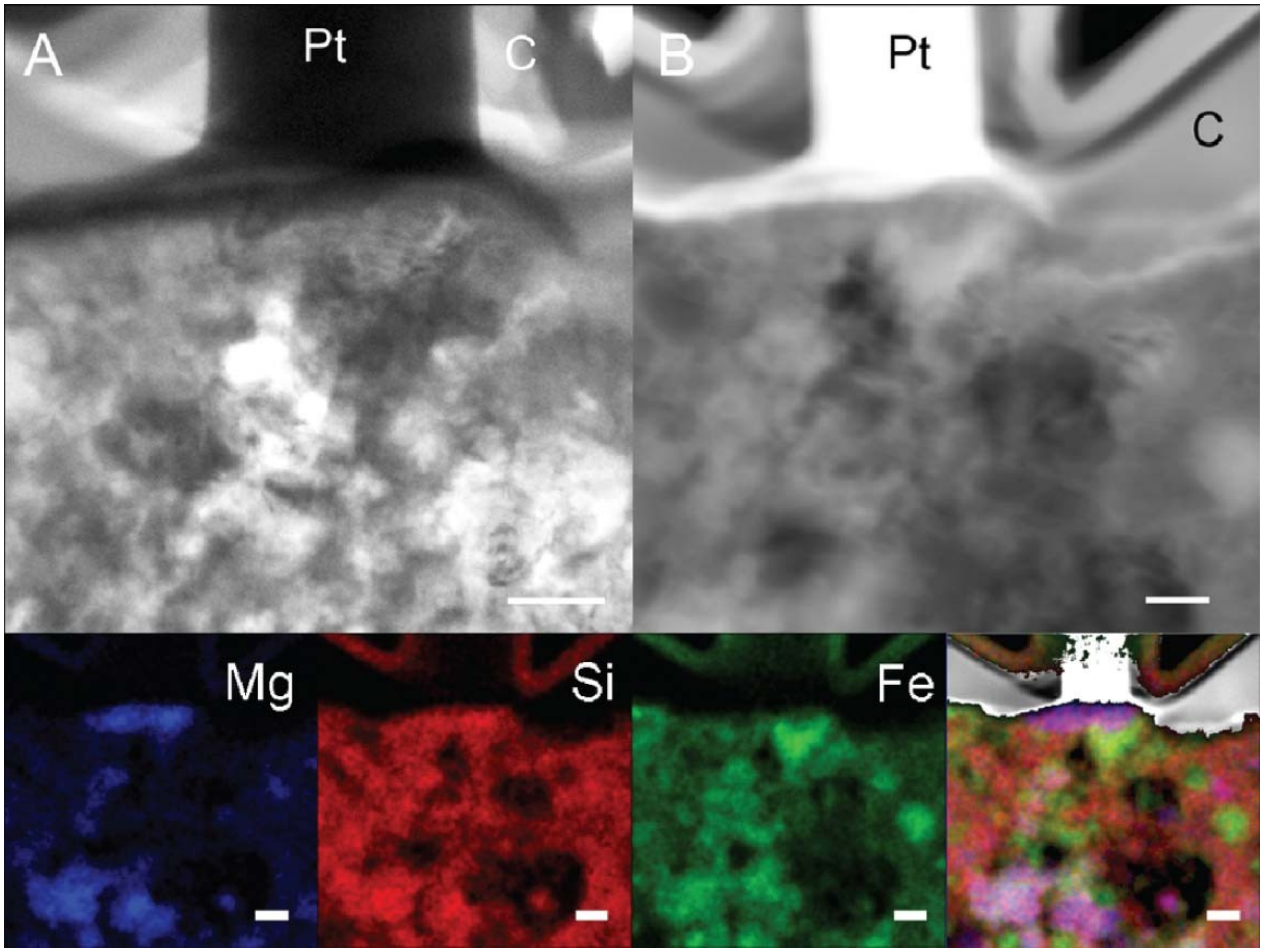

Figure 6. Bright field TEM (A), HAADF (B) and STEM-EDX (bottom) maps of the FIB cross-section of grain 139a. The EDX maps are shown as net counts at the K edge of each element. The boundary of the presolar grain is clearly visible in the Mg image. The labels Pt and C refer to the FIB-deposited platinum and carbon masks. (scale bars = 100 nm)

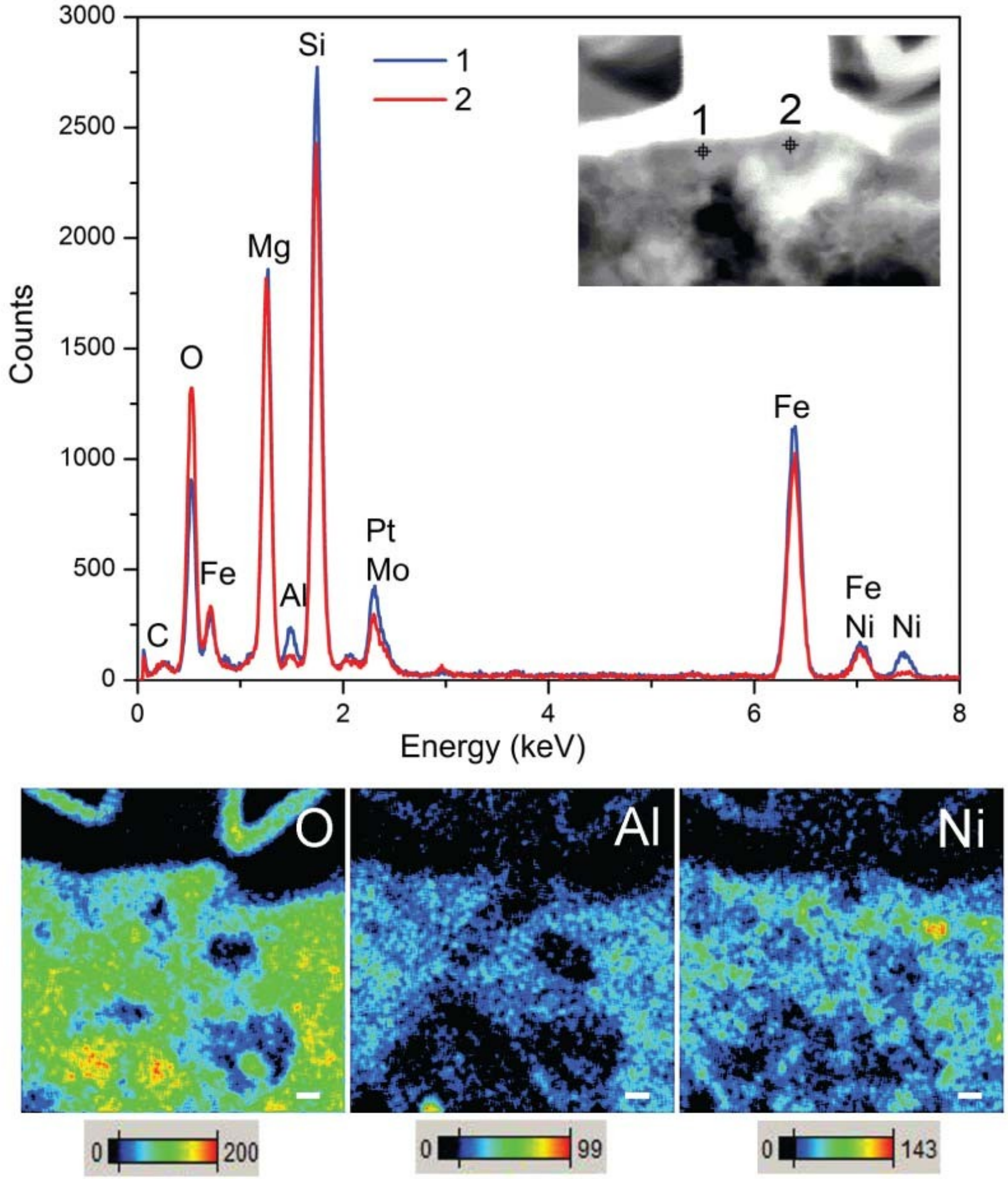

Figure 7. STEM-EDX point spectra and maps from the FIB cross-section of grain 139a. The left side (1) of the grain is distinctly lower in O and higher in Al and Ni than the right side (2). (scale bars = 100 nm)

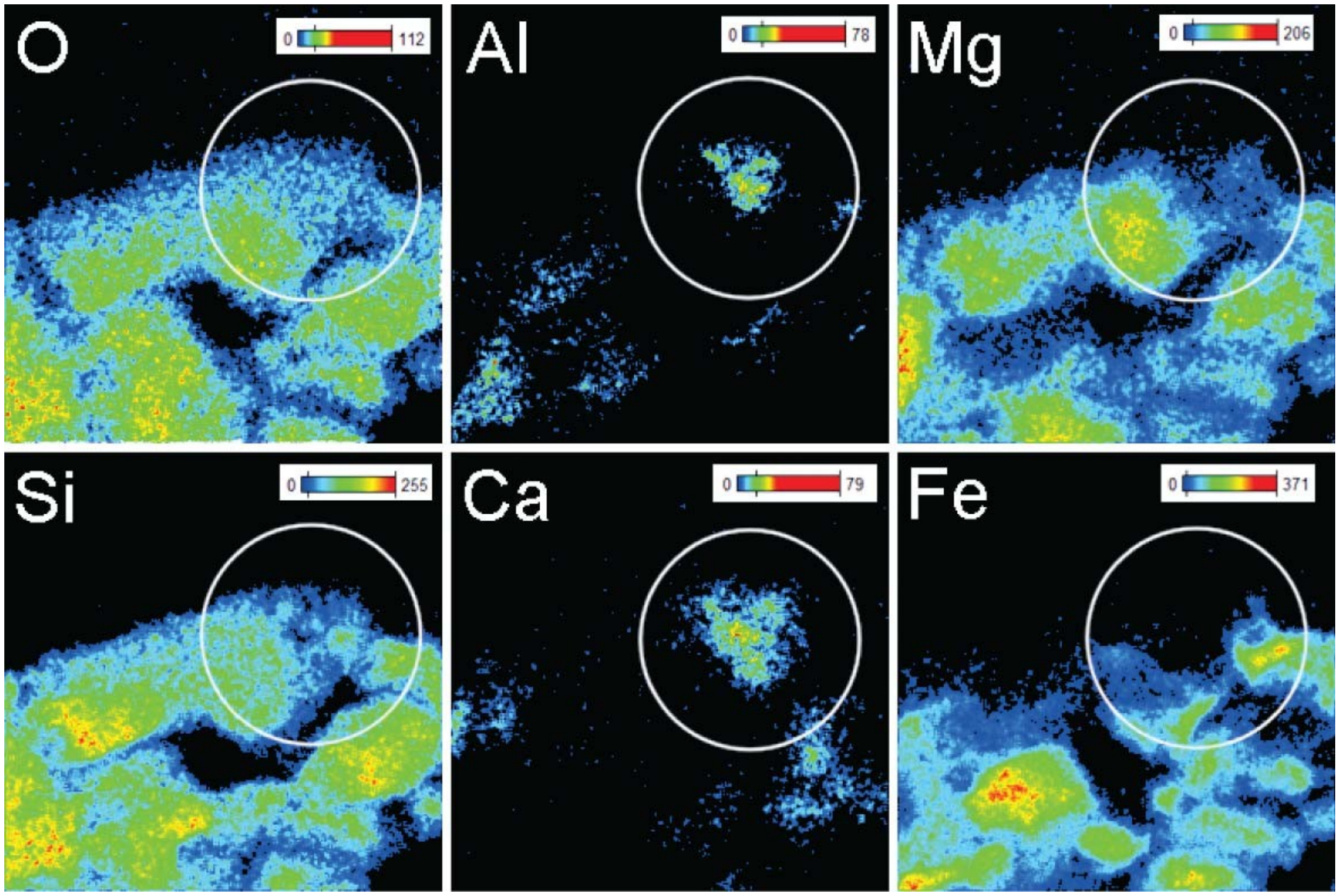

Figure 8. STEM-EDX maps of the FIB cross-section of grain 166a. The 600-nm white circle, centered on grain 166a, is shown as an aid to comparison with Auger elemental maps of the grain surface in Figure 4. In the cross-section it is clear that this circle includes a portion of a Mg-rich silicate grain to the left of the presolar grain. Mg is distributed heterogeneously across grain 166a, but does not appear to form a distinct rim. Si, Al, and Ca are also heterogeneously distributed at a ~30 nm scale.

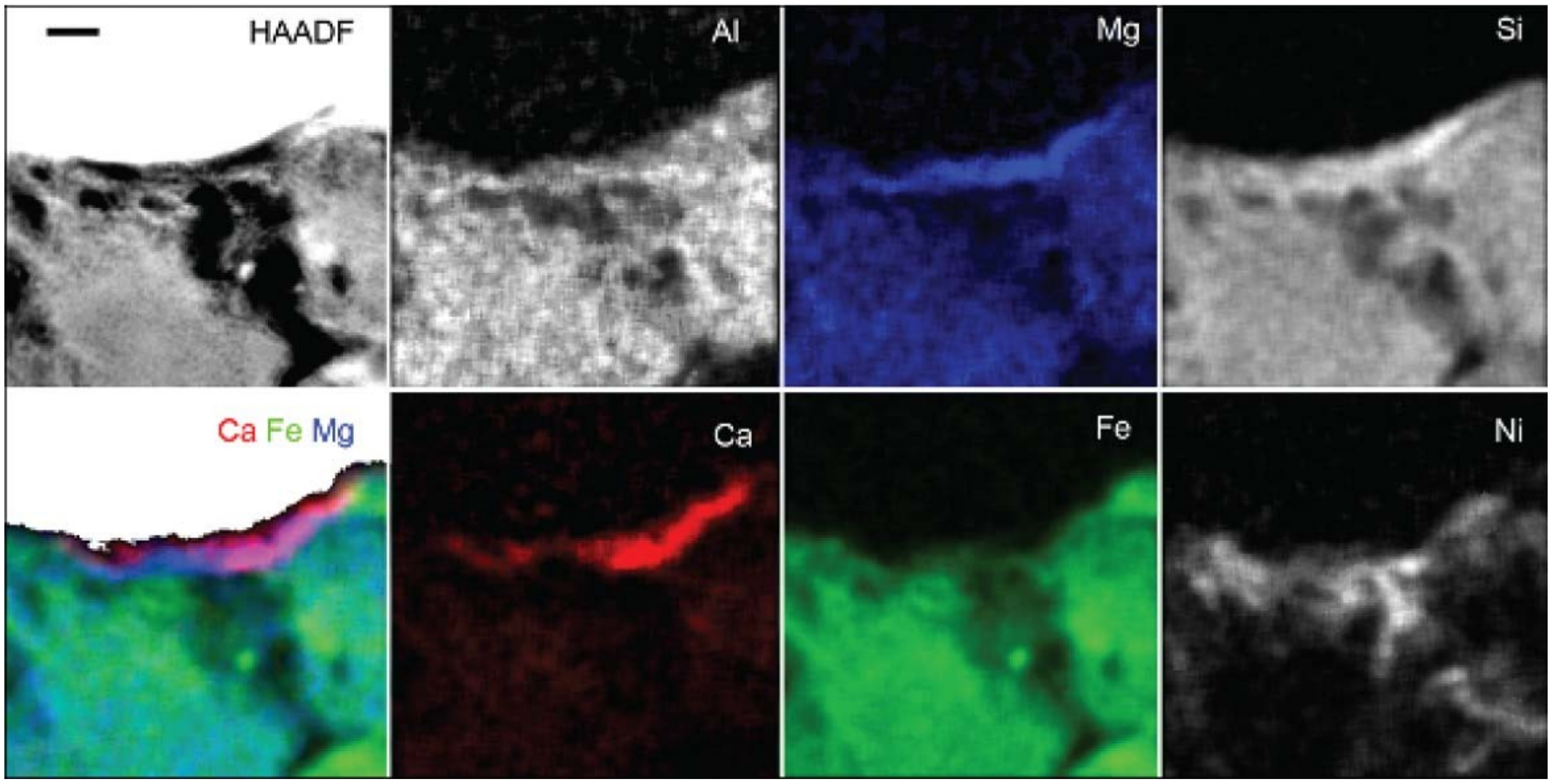

Figure 9. STEM-EDX maps and high-angle annular dark field (HAADF) image of the FIB cross-section of grain 65a. The elemental maps are shown as K-edge net counts with 9 × 9 pixel averaging. The white region of the HAADF and composite images is the Pt mask. The Ca,Mg-rich presolar grain is 500 nm across, but extends only ~30 nm below the Pt mask. (scale bar =100 nm)

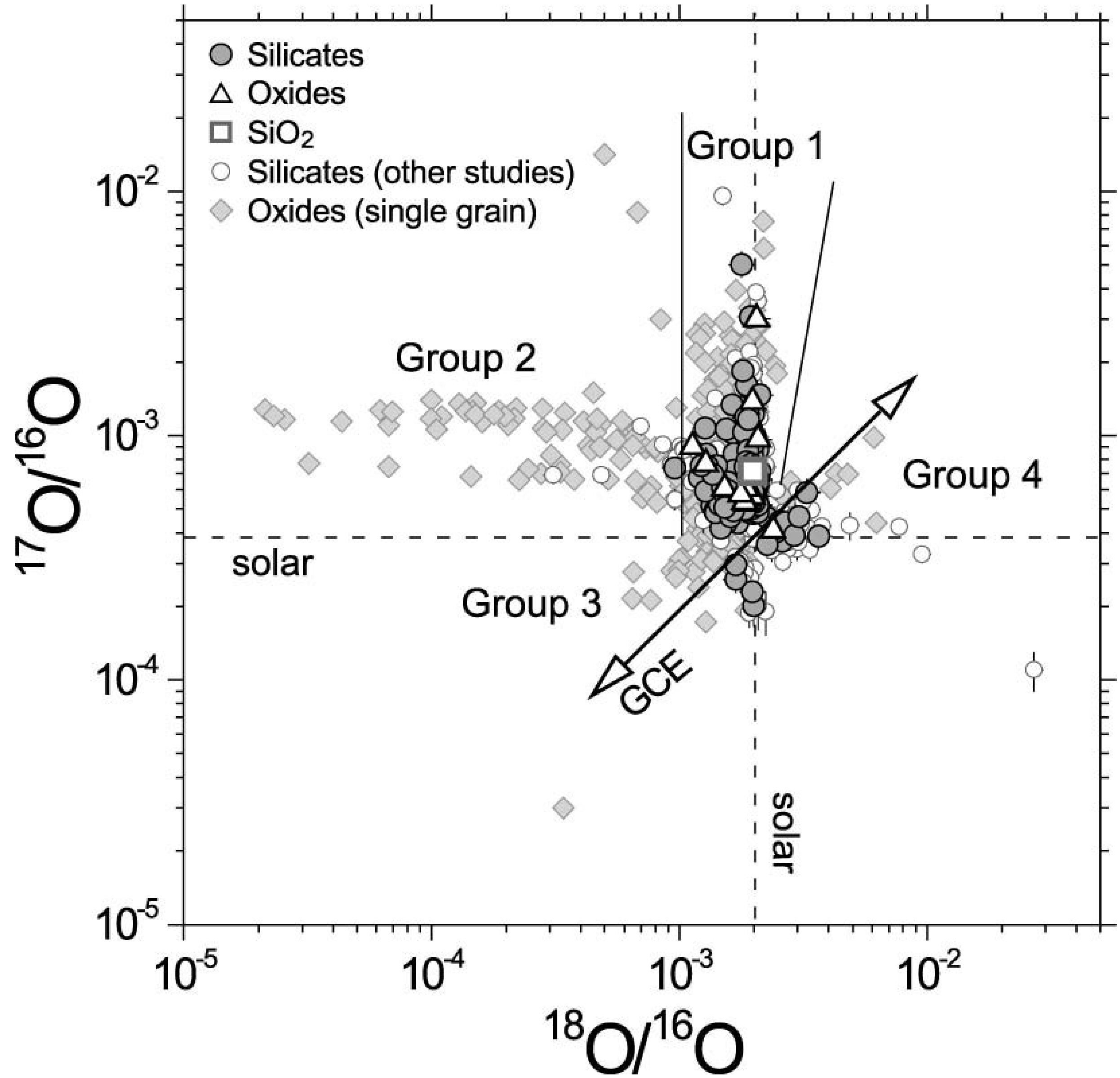

Figure 10. Oxygen isotopic compositions of presolar grains identified in this study compared to presolar silicates identified in other studies (Bland et al. 2007; Floss & Stadermann 2009; Floss et al. 2006; Messenger, Keller, & Lauretta 2005; Messenger et al. 2003a; Mostefaoui & Hoppe 2004; Nguyen et al. 2007b; Nguyen & Zinner 2004; Vollmer et al. 2009b; Yada et al. 2008) also by NanoSIMS raster ion imaging, and to presolar oxides identified in previous studies by single grain measurement. The four oxide group delineations (Nittler et al. 1997b) are shown, as well as the Galactic chemical evolution (GCE) trend. The dashed lines indicate solar compositions. All presolar silicates appear to have a more limited compositional range than presolar oxides due to the isotopic dilution effect. Most of the grain compositions can be explained by GCE and nucleosynthetic processes, but some of them are likely signatures of SN or novae.

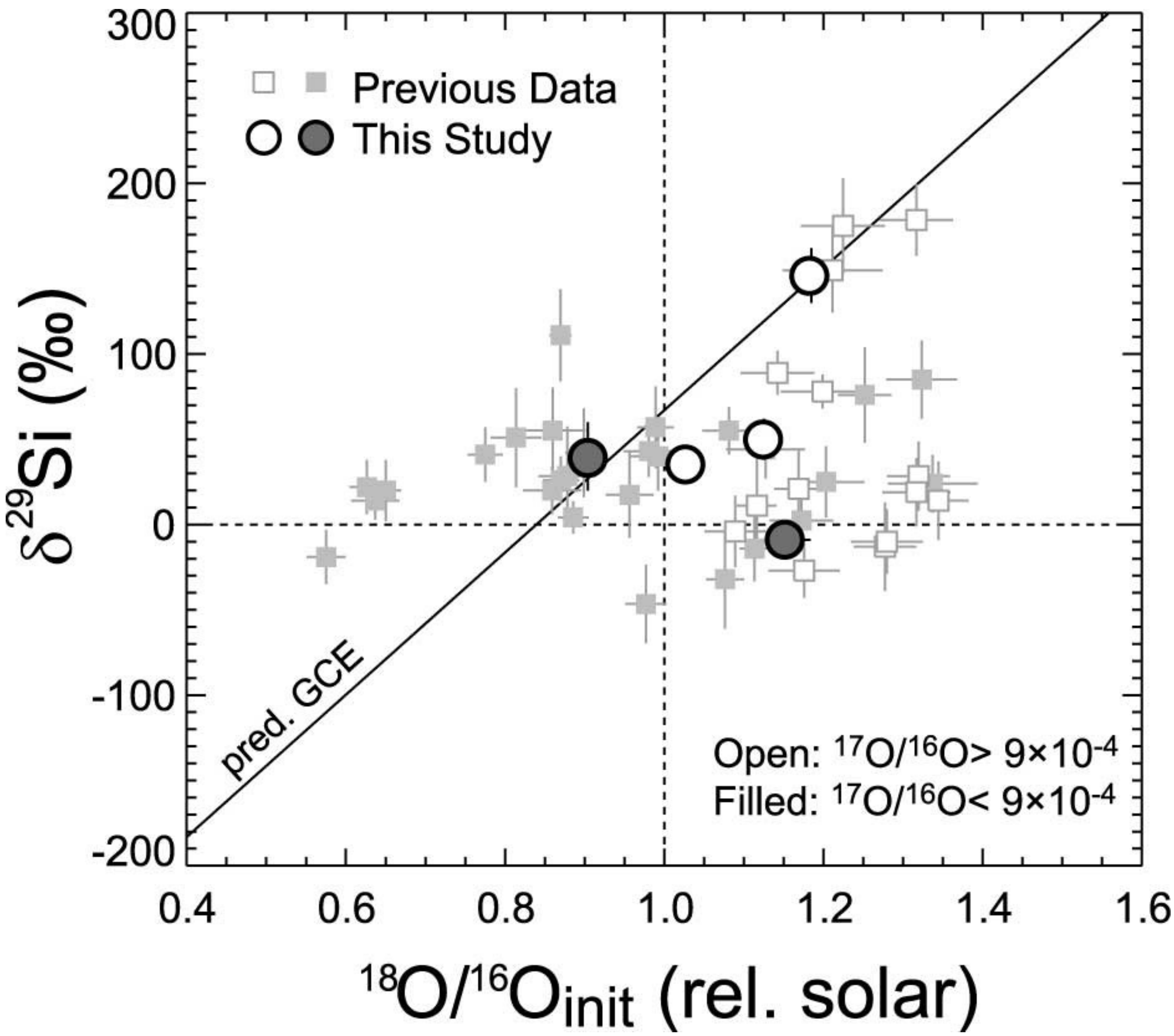

Figure 11. $\delta^{29}$Si of presolar silicate grains from this study and others (Busemann et al. 2009; Mostefaoui & Hoppe 2004; Nguyen et al. 2007b; Vollmer, Hoppe, & Brenker 2008) plotted against the inferred initial $^{18}O/^{16}O$ ratios relative to solar. The solid line indicates the predicted trend for Galactic chemical evolution, based on relations between Si and Ti isotopes in presolar SiC grains and Ti and O isotopes in presolar $Al_2O_3$ grains. The data, which suffer from isotopic dilution, do not appear to follow the predicted trend. The dashed lines indicate solar isotopic compositions.

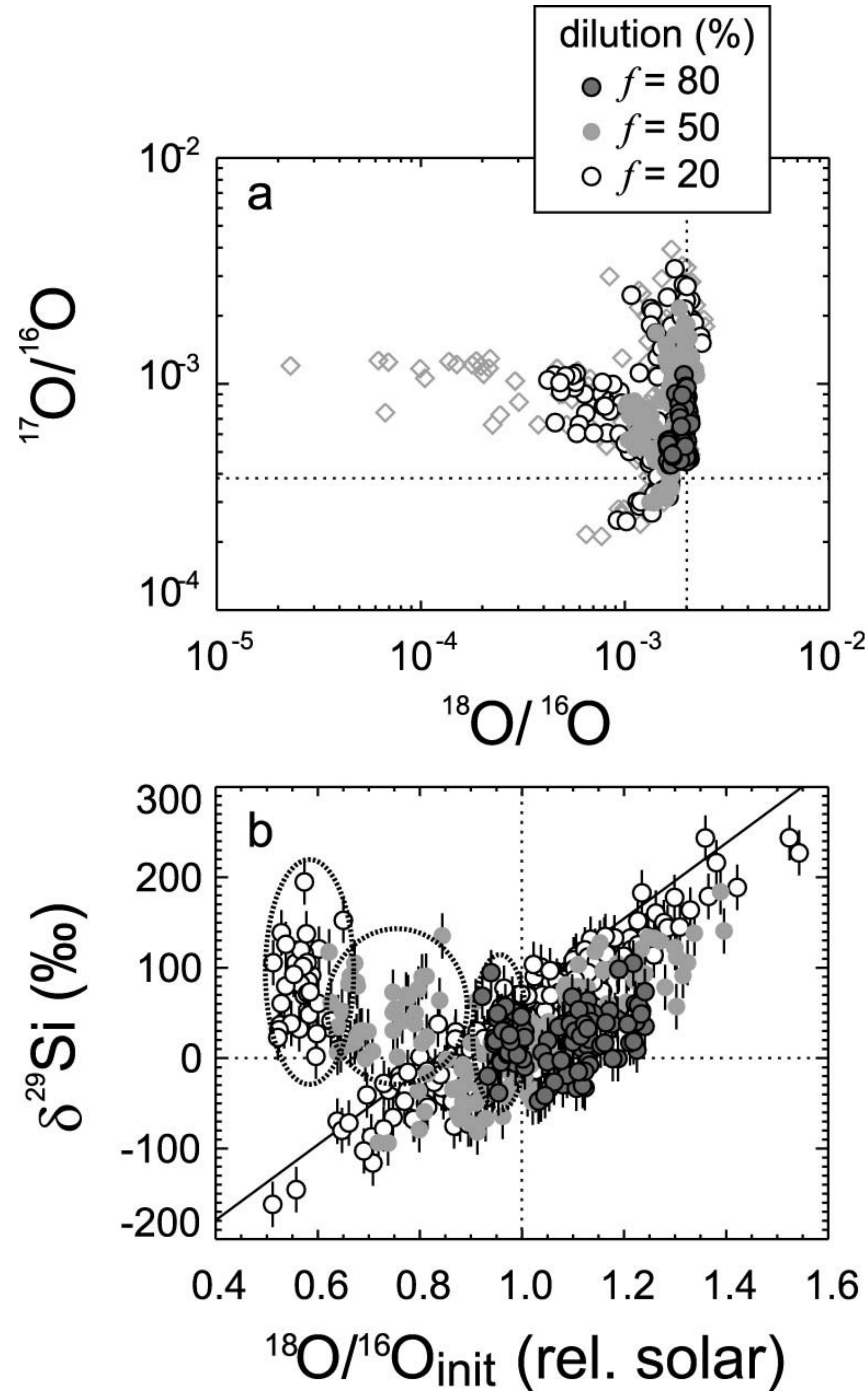

Figure 12. Calculations illustrating the effect of isotopic dilution on measured O isotopic compositions (a) and on the relation between $\delta^{29}$Si and inferred initial parent stellar metallicity, given as initial $^{18}O/^{16}O$ relative to solar system value (b). The isotopic compositions of Group 1-3 presolar oxides identified by individual grain analysis (diamonds) were taken as the starting values. The solid line indicates the same assumed relationship between $\delta^{29}$Si and initial $^{18}O/^{16}O$ prior to isotope dilution as in Figure 11. Ellipses in (b) indicate Group 2 grains. Dashed lines indicate solar compositions. See text for details.

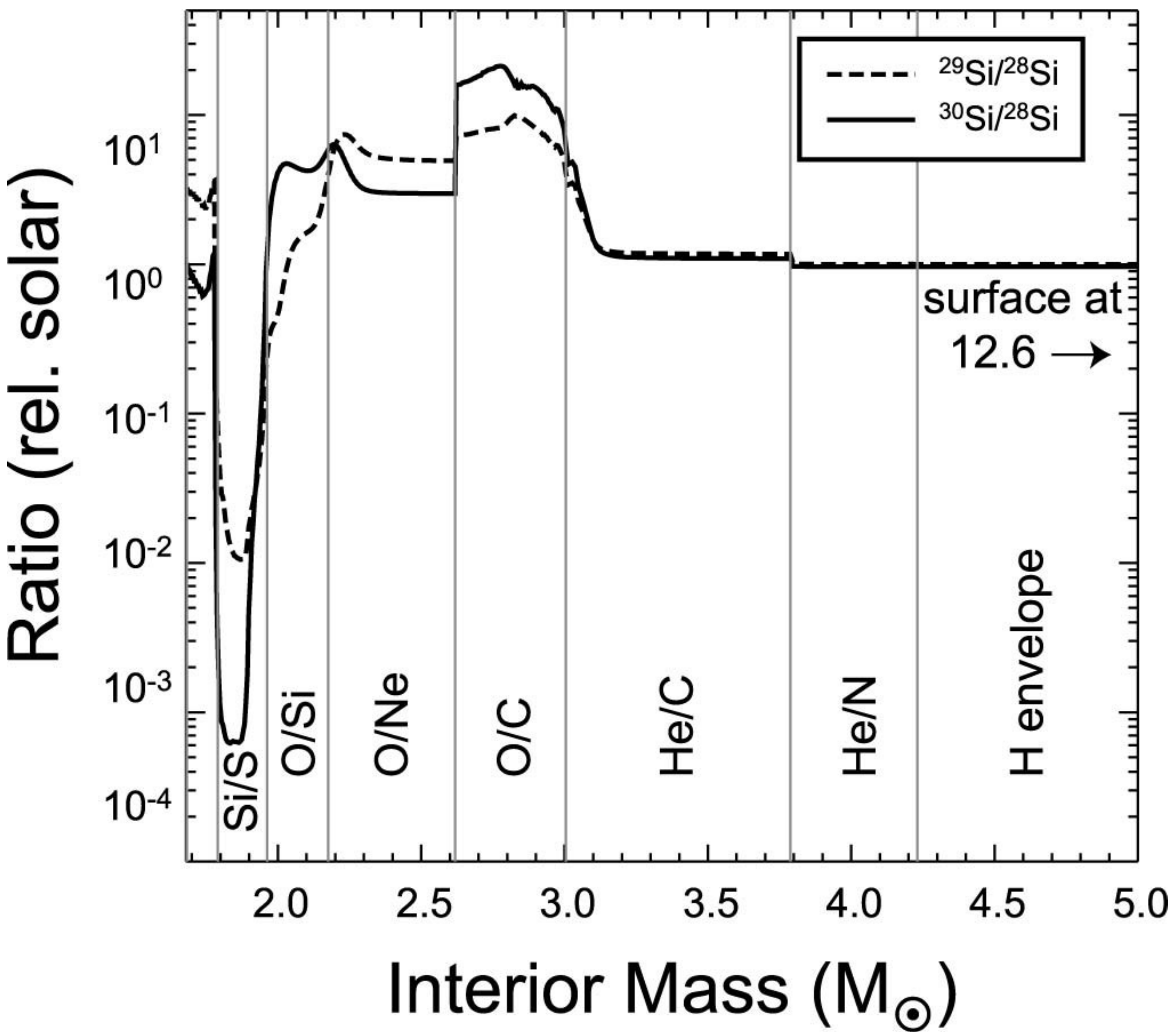

Figure 13. Predictions for Si isotopic ratios in the different zones of a 15$M_{\odot}$ supernova from the model of Rauscher et al. (2002). Larger variations are observed for the inner O- and Si-rich zones.

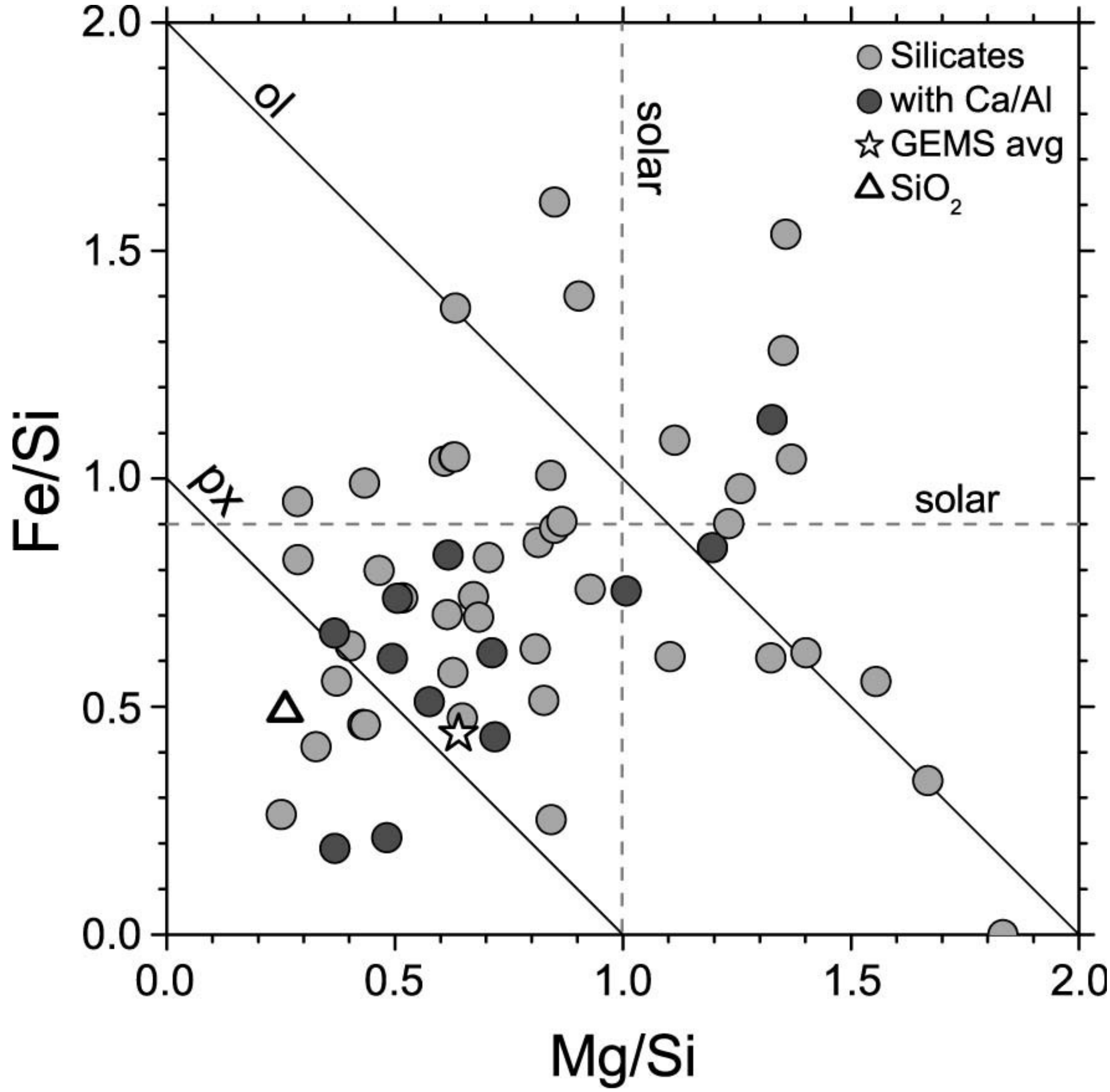

Figure 14. Mg and Fe contents of presolar silicate and silica grains, relative to Si, determined by Auger spectroscopy. Also shown is the average composition of GEMS grains (Keller & Messenger 2004). The vast majority of presolar silicates are not stoichiometric pyroxene or olivine (solid lines), and many grains have sub-solar Fe/Si and Mg/Si compositions, similar to GEMS in IDPs.

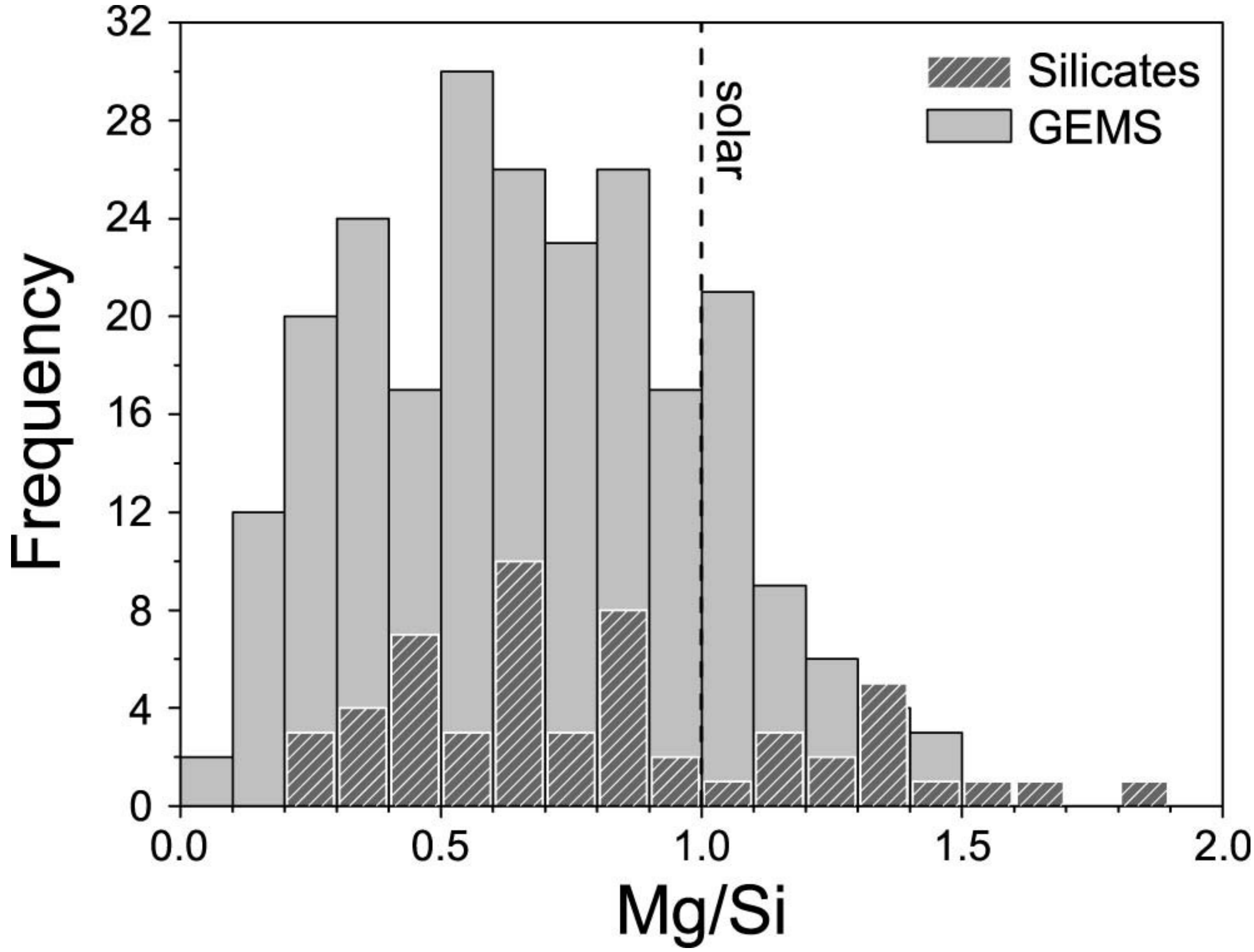

Figure 15. Histogram of Mg/Si ratios in presolar silicates from this study, as determined by Auger spectroscopy, and in GEMS from IDPs (Keller & Messenger 2004). The two distributions are similar, with most of the grains having sub-solar Mg/Si ratios.

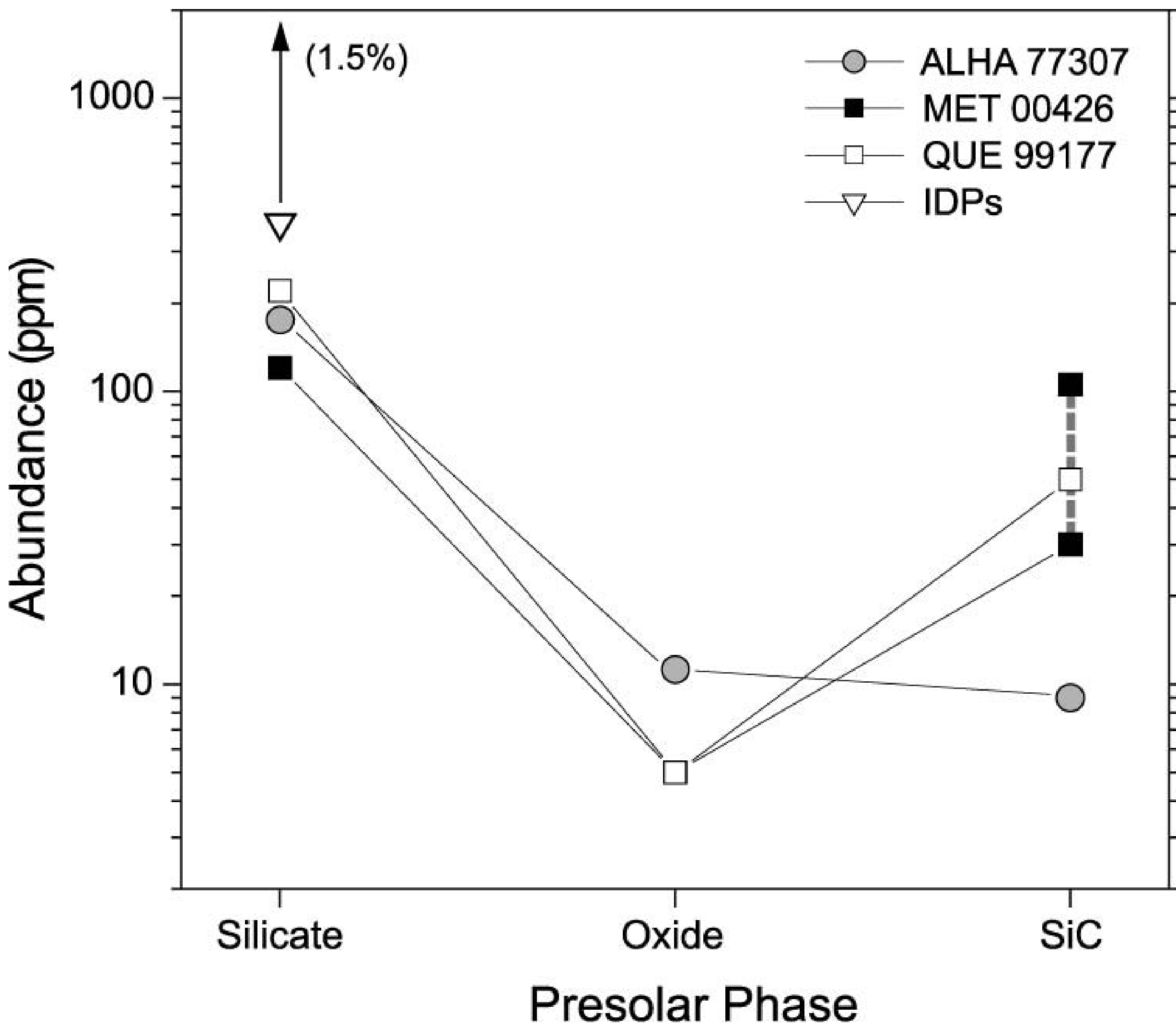

Figure 16. Matrix normalized abundances, in parts per million, of presolar phases in ALHA 77307, MET 00426 (Floss & Stadermann 2008), QUE 99177 (Floss & Stadermann 2008), and in IDPs (Floss et al. 2006; Nguyen, Busemann, & Nittler 2007a). No corrections for detection efficiency were applied. The SiC abundance in ALHA 77307 was taken from literature (Davidson et al. 2009; Huss et al. 2003). Only one presolar oxide and one presolar SiC have been identified in IDPs (Stadermann et al. 2006b). The lack of a systematic variance among the samples suggests nebular alteration alone cannot explain these presolar abundances.

Table 1. Oxygen and Silicon isotopic compositions, diameters, and phases of all presolar grains identified in ALHA 77307.

| Grain | $^{17}O/^{16}O$ ($\times10^{-4}$) | $^{18}O/^{16}O$ ($\times10^{-3}$) | $\delta^{17}O/^{16}O$ | $\delta^{18}O/^{16}O$ | $\delta^{29}Si/^{28}Si$ | $\delta^{30}Si/^{28}Si$ | Size (nm) | Phase (NanoSIMS) |
|---|---|---|---|---|---|---|---|---|
| AH-84 | 9.02 ± 0.66 | 1.13 ± 0.09 | 1355 ± 174 | -434 ± 44 | | | 235 | Mg-poor oxide |
| AH-100a | 7.33 ± 0.62 | 1.99 ± 0.09 | 914 ± 162 | -6 ± 43 | | | 215 | oxide |
| AH-101a | 13.75 ± 0.57 | 1.97 ± 0.07 | 2590 ± 149 | -20 ± 35 | | | 280 | Al-rich oxide |
| AH-102 | 9.62 ± 0.63 | 2.07 ± 0.07 | 1512 ± 164 | 34 ± 34 | | | 280 | oxide |
| AH-107b | 5.55 ± 0.44 | 1.87 ± 0.07 | 448 ± 114 | -70 ± 37 | | | 250 | oxide |
| AH-111a | 30.12 ± 0.78 | 2.05 ± 0.06 | 6863 ± 205 | 23 ± 30 | | | 305 | Al-rich oxide |
| AH-114 | 7.67 ± 0.39 | 1.28 ± 0.05 | 1002 ± 102 | -362 ± 25 | | | 330 | Al-rich oxide |
| AH-127 | 5.64 ± 0.39 | 1.97 ± 0.06 | 472 ± 101 | -19 ± 32 | | | 265 | oxide |
| AH-129[#] | 6.79 ± 0.47 | 1.94 ± 0.08 | 773 ± 121 | -31 ± 39 | | | 265 | oxide |
| AH-144 | 6.14 ± 0.45 | 1.89 ± 0.07 | 603 ± 118 | -58 ± 35 | | | 280 | oxide |
| AH-147b | 5.27 ± 0.28 | 1.87 ± 0.04 | 376 ± 73 | -68 ± 22 | | | 340 | oxide |
| AH-90[&] | 5.88 ± 0.33 | 2.03 ± 0.06 | 535 ± 86 | 14 ± 30 | -13 ± 69 | -175 ± 91 | 330 | Mg-rich oxide |
| AH-132b[&] | 5.01 ± 0.26 | 1.59 ± 0.05 | 307 ± 67 | -208 ± 24 | -17 ± 46 | -16 ± 57 | 350 | oxide |
| AH-148b[&] | 7.81 ± 0.46 | 1.89 ± 0.07 | 1038 ± 120 | -58 ± 36 | 71 ± 76 | -50 ± 82 | 265 | oxide |
| AH-148d[&] | 4.64 ± 0.37 | 1.61 ± 0.07 | 212 ± 97 | -198 ± 33 | -3 ± 60 | 0 ± 73 | 250 | oxide |
| AH-151[&] | 14.75 ± 0.57 | 2.11 ± 0.09 | 2850 ± 148 | 52 ± 42 | -2 ± 50 | 65 ± 63 | 355 | oxide |
| AH-153[&] | 7.80 ± 0.51 | 1.87 ± 0.07 | 1037 ± 134 | -68 ± 33 | -74 ± 57 | 55 ± 75 | 305 | oxide |
| AH-33a | 7.14 ± 0.39 | 1.98 ± 0.08 | 863 ± 101 | -14 ± 38 | 118 ± 47 | 3 ± 53 | 175 | silicate |
| AH-33b[◊] | 4.58 ± 0.22 | 2.27 ± 0.07 | 195 ± 58 | 131 ± 32 | -6 ± 35 | -15 ± 42 | 215 | silicate |
| AH-36 | 7.18 ± 0.45 | 1.95 ± 0.06 | 873 ± 118 | -29 ± 32 | 113 ± 55 | 35 ± 69 | 235 | silicate |
| AH-42 | 2.58 ± 0.30 | 1.69 ± 0.08 | -325 ± 78 | -156 ± 39 | 14 ± 54 | -104 ± 62 | 195 | silicate |
| AH-43 | 7.55 ± 0.51 | 1.60 ± 0.07 | 972 ± 133 | -200 ± 37 | 63 ± 57 | 1 ± 68 | 215 | silicate |
| AH-46 | 6.73 ± 0.32 | 1.38 ± 0.05 | 758 ± 82 | -310 ± 25 | | | 280 | silicate |
| AH-47 | 6.05 ± 0.48 | 1.38 ± 0.07 | 578 ± 126 | -310 ± 34 | | | 175 | silicate |
| AH-48a | 6.43 ± 0.29 | 2.00 ± 0.06 | 679 ± 75 | -5 ± 30 | | | 265 | silicate |
| AH-48b | 4.84 ± 0.21 | 1.95 ± 0.04 | 263 ± 56 | -27 ± 21 | | | 290 | silicate |
| AH-48c | 2.95 ± 0.29 | 1.69 ± 0.05 | -230 ± 76 | -156 ± 26 | | | 250 | silicate |
| AH-49a | 4.38 ± 0.23 | 1.72 ± 0.06 | 143 ± 59 | -145 ± 31 | | | 265 | silicate |
| AH-50a | 6.36 ± 0.55 | 1.32 ± 0.07 | 660 ± 144 | -344 ± 35 | | | 175 | silicate |
| AH-50b | 5.65 ± 0.32 | 2.05 ± 0.06 | 475 ± 84 | 23 ± 30 | | | 215 | silicate |
| AH-56 | 5.55 ± 0.25 | 2.02 ± 0.05 | 450 ± 65 | 5 ± 24 | | | 365 | silicate |
| AH-59 | 6.98 ± 0.36 | 2.04 ± 0.08 | 822 ± 93 | 16 ± 37 | | | 265 | silicate |
| AH-60 | 6.86 ± 0.33 | 1.67 ± 0.06 | 791 ± 87 | -168 ± 30 | | | 290 | silicate |
| AH-63 | 6.76 ± 0.53 | 1.92 ± 0.10 | 764 ± 138 | -41 ± 52 | | | 250 | silicate |
| AH-65a[%] | 8.36 ± 0.27 | 1.29 ± 0.02 | 1181 ± 69 | -357 ± 12 | | | 615 | silicate |
| AH-65b | 4.85 ± 0.24 | 1.91 ± 0.04 | 266 ± 62 | -49 ± 18 | | | 415 | silicate |
| AH-70 | 8.76 ± 0.37 | 2.07 ± 0.06 | 1288 ± 96 | 34 ± 28 | | | 265 | Mg-rich silicate |
| AH-72 | 6.95 ± 0.52 | 2.07 ± 0.08 | 813 ± 136 | 34 ± 41 | | | 280 | silicate |
| AH-73a | 13.07 ± 0.43 | 1.76 ± 0.05 | 2412 ± 113 | -124 ± 26 | | | 320 | silicate |
| AH-73b | 7.26 ± 0.34 | 2.01 ± 0.06 | 895 ± 89 | 0 ± 28 | | | 295 | silicate |
| AH-73c | 3.69 ± 0.27 | 2.60 ± 0.07 | -38 ±71 | 297 ± 33 | | | 265 | silicate |
| AH-74 | 5.93 ± 0.44 | 1.90 ± 0.07 | 548 ± 115 | -51 ± 37 | | | 250 | Mg-rich silicate |
| AH-75 | 6.54 ± 0.50 | 1.73 ± 0.06 | 707 ± 130 | -139 ± 31 | | | 265 | silicate |
| AH-83a | 6.18 ± 0.33 | 1.81 ± 0.05 | 612 ± 87 | -99 ± 27 | 22 ± 41 | 1 ± 50 | 305 | silicate |
| AH-83b | 5.16 ± 0.47 | 1.35 ± 0.07 | 346 ± 122 | -326 ± 37 | -74 ± 66 | 125 ± 79 | 195 | silicate |
| AH-85a | 10.66 ± 0.83 | 1.27 ± 0.10 | 1783 ± 217 | -365 ± 49 | 153 ± 72 | 4 ± 92 | 175 | silicate |
| AH-85b | 6.21 ± 0.33 | 1.98 ± 0.06 | 622 ± 85 | -14 ± 29 | 29 ± 44 | 100 ± 56 | 330 | silicate |
| AH-85c | 5.79 ± 0.43 | 1.98 ± 0.08 | 511 ± 113 | -13 ± 41 | 11 ± 62 | 54 ± 97 | 235 | silicate |
| AH-85d | 4.90 ± 0.47 | 1.39 ± 0.11 | 278 ± 122 | -308 ± 56 | 225 ± 94 | 51 ± 121 | 195 | silicate |
| AH-86 | 6.70 ± 0.58 | 1.54 ± 0.08 | 748 ± 151 | -231 ± 42 | -113 ± 68 | -99 ± 94 | 215 | silicate |
| AH-87a | 7.14 ± 0.52 | 2.11 ± 0.09 | 864 ± 136 | 53 ± 46 | 14 ± 76 | 5 ± 106 | 215 | silicate |
| AH-87b | 4.03 ± 0.46 | 2.45 ± 0.10 | 53 ± 120 | 224 ± 52 | -15 ± 86 | -69 ± 103 | 195 | silicate |
| AH-88a | 7.40 ± 0.44 | 2.04 ± 0.08 | 933 ± 114 | 16 ± 39 | 75 ± 57 | 150 ± 86 | 265 | silicate |
| AH-88b | 12.22 ± 0.60 | 1.99 ± 0.07 | 2190 ± 156 | -10 ± 36 | 27 ± 80 | -122 ± 74 | 280 | silicate |
| AH-89 | 6.56 ± 0.58 | 1.66 ± 0.08 | 712 ± 151 | -170 ± 39 | -88 ± 68 | -185 ± 79 | 250 | Mg-poor silicate |
| AH-96a | 16.07 ± 0.61 | 1.87 ± 0.06 | 3194 ± 159 | -66 ± 32 | 120 ± 64 | 137 ± 78 | 305 | silicate |
| AH-96b | 6.45 ± 0.46 | 1.89 ± 0.08 | 684 ± 120 | -56 ± 40 | 17 ± 90 | -105 ± 107 | 235 | silicate |
| AH-98a | 6.25 ± 0.55 | 2.07 ± 0.10 | 632 ± 144 | 32 ± 50 | 89 ± 87 | 237 ± 116 | 195 | silicate |
| AH-98b | 2.01 ± 0.34 | 1.99 ± 0.11 | -476 ± 90 | -6 ± 55 | 90 ± 88 | 284 ± 146 | 235 | silicate |
| AH-100b | 6.20 ± 0.59 | 1.90 ± 0.11 | 618 ± 155 | -51 ± 55 | -42 ± 95 | -181 ± 98 | 215 | silicate |
| AH-101b | 5.64 ± 0.49 | 1.64 ± 0.09 | 473 ± 128 | -183 ± 47 | -61 ± 94 | -78 ± 97 | 250 | silicate |
| AH-103 | 4.39 ± 0.40 | 1.55 ± 0.07 | 146 ± 104 | -225 ± 37 | 153 ± 104 | 42 ± 114 | 265 | silicate |
| AH-104a | 6.72 ± 0.37 | 1.20 ± 0.05 | 755 ± 96 | -404 ± 27 | 52 ± 65 | 7 ± 72 | 340 | Al-rich silicate |
| AH-104b | 6.13 ± 0.48 | 1.73 ± 0.11 | 601 ± 126 | -137 ± 53 | 62 ± 92 | 15 ± 86 | 250 | silicate |
| AH-105 | 6.75 ± 0.38 | 1.95 ± 0.06 | 763 ± 99 | -28 ± 29 | -14 ± 67 | -56 ± 73 | 320 | silicate |
| AH-106a | 50.11 ± 2.16 | 1.78 ± 0.07 | 12081 ± 563 | -114 ± 33 | 15 ± 59 | 80 ± 67 | 305 | silicate |
| AH-106b | 6.61 ± 0.49 | 1.88 ± 0.10 | 725 ± 127 | -65 ± 50 | -50 ± 86 | 198 ± 97 | 250 | silicate |
| AH-107a | 10.21 ± 0.48 | 1.64 ± 0.05 | 1666 ± 126 | -180 ± 25 | -26 ± 46 | 88 ± 59 | 365 | silicate |
| AH-111b | 8.56 ± 0.49 | 1.97 ± 0.06 | 1235 ± 128 | -18 ± 30 | -25 ± 55 | -59 ± 73 | 305 | silicate |
| AH-111c | 7.47 ± 0.46 | 1.81± 0.09 | 950 ± 119 | -99 ± 43 | -34 ± 81 | 8 ± 99 | 265 | silicate |
| AH-112 | 8.42 ± 0.42 | 1.66 ± 0.06 | 1198 ± 109 | -170 ± 30 | 70 ± 56 | 149 ± 74 | 320 | silicate |
| AH-115 | 8.13 ± 0.36 | 1.98 ± 0.07 | 1122 ± 94 | -14 ± 33 | 63 ± 51 | 117 ± 65 | 385 | silicate |
| AH-116 | 2.30 ± 0.34 | 1.97 ± 0.10 | -399 ± 88 | -19 ± 49 | -15 ± 95 | -22 ± 96 | 250 | silicate |
| AH-117a | 6.67 ± 0.52 | 1.87 ± 0.07 | 741 ± 136 | -66 ± 36 | -23 ± 88 | 101 ± 88 | 305 | silicate |
| AH-118 | 6.15 ± 0.42 | 2.03 ± 0.10 | 604 ± 110 | 13 ± 50 | 143 ± 86 | 134 ± 105 | 250 | silicate |
| AH-119a | 9.92 ± 0.54 | 2.02 ± 0.07 | 1589 ± 142 | 7 ± 36 | 53 ± 85 | 66 ± 86 | 280 | silicate |
| AH-119b | 6.07 ± 0.50 | 1.64 ± 0.10 | 585 ± 130 | -180 ± 48 | 79 ± 82 | -65 ± 93 | 215 | silicate |
| AH-121 | 3.88 ± 0.27 | 3.66 ± 0.08 | 14 ± 71 | 823 ± 39 | 2 ± 67 | -72 ± 69 | 330 | silicate |
| AH-122 | 10.59 ± 0.57 | 1.96 ± 0.07 | 1763 ± 150 | -22 ± 34 | 79 ± 75 | 156 ± 94 | 290 | silicate |
| AH-123 | 4.17 ± 0.32 | 1.47 ± 0.07 | 88 ± 83 | -265 ± 34 | 126 ± 67 | 65 ± 69 | 330 | silicate |
| AH-124a | 18.44 ± 1.00 | 1.80 ± 0.11 | 3813 ± 262 | -103 ± 57 | 29 ± 80 | 116 ± 94 | 320 | silicate |
| AH-124b | 6.33 ± 0.38 | 1.74 ± 0.06 | 652 ± 99 | -132 ± 31 | 51 ± 60 | -40 ± 69 | 280 | silicate |
| AH-126 | 10.54 ± 0.51 | 1.94 ± 0.06 | 1751 ± 134 | -32 ± 30 | 123 ± 68 | 13 ± 95 | 320 | silicate |
| AH-130 | 7.21 ± 0.33 | 1.65 ± 0.05 | 882 ± 87 | -176 ± 25 | -95 ± 38 | 10 ± 50 | 330 | silicate |
| AH-132a | 30.64 ± 0.87 | 1.93 ± 0.05 | 6999 ± 228 | -38 ± 25 | 85 ± 42 | 98 ± 52 | 350 | silicate |
| AH-132c◈ | 9.21 ± 0.67 | 1.99 ± 0.06 | 1403 ± 176 | -7 ± 32 | -37 ± 52 | 73 ± 74 | 350 | silicate |
| AH-132d◈ | 7.66 ± 0.62 | 1.94 ± 0.08 | 998 ± 162 | -30 ± 39 | 64 ± 93 | -36 ± 93 | 330 | silicate |
| AH-133 | 10.63 ± 0.50 | 1.55 ± 0.06 | 1774 ± 130 | -227 ± 29 | 23 ± 51 | 54 ± 62 | 470 | silicate |
| AH-136 | 5.20 ± 0.38 | 1.51 ± 0.06 | 357 ± 99 | -245 ± 30 | -6 ± 47 | 4 ± 58 | 330 | silicate |
| AH-137a◈ | 5.61 ± 0.32 | 1.61 ± 0.05 | 465 ± 85 | -196 ± 27 | 0 ± 47 | 0 ± 61 | 310 | silicate |
| AH-137b◈ | 5.19 ± 0.29 | 2.08 ± 0.06 | 354 ± 76 | 39 ± 29 | 48 ± 49 | 42 ± 59 | 310 | silicate |
| AH-138a | 7.72 ± 0.25 | 1.92 ± 0.04 | 1015 ± 66 | -42 ± 21 | 112 ± 45 | -2 ± 42 | 375 | silicate |
| AH-138b | 5.44 ± 0.34 | 1.93 ± 0.06 | 421 ± 88 | -37 ± 28 | 72 ± 45 | -37 ± 59 | 265 | silicate |
| AH-138c | 4.44 ± 0.25 | 2.53± 0.06 | 159 ± 64 | 260 ± 29 | -63 ± 60 | 29 ± 56 | 305 | silicate |
| AH-138d | 4.16 ± 0.26 | 2.45 ± 0.07 | 85 ± 67 | 222 ± 32 | -44 ± 47 | 55 ± 53 | 250 | silicate |
| AH-139a[%] | 10.35 ± 0.15 | 1.82 ± 0.02 | 1701 ± 40 | -95 ± 9 | 50 ± 12 | 30 ± 14 | 520 | silicate |
| AH-139b | 5.44 ± 0.24 | 1.75 ± 0.04 | 420 ± 63 | -126 ± 19 | 58 ± 39 | 20 ± 47 | 350 | silicate |
| AH-139c | 4.84 ± 0.28 | 1.40 ± 0.05 | 264 ± 72 | -301 ± 23 | 37 ± 44 | 24 ± 55 | 290 | silicate |
| AH-140 | 6.56 ± 0.32 | 1.45 ± 0.04 | 713 ± 84 | -275 ± 21 | 4 ± 40 | 10 ± 50 | 320 | silicate |
| AH-142 | 5.08 ± 0.30 | 1.64 ± 0.08 | 326 ± 79 | -181 ± 41 | -10 ± 57 | 66 ± 57 | 250 | silicate |
| AH-145 | 7.37 ± 0.52 | 0.96 ± 0.05 | 925 ± 135 | -522 ± 26 | 64 ± 45 | -66 ± 54 | 290 | silicate |
| AH-147a | 6.30 ± 0.26 | 1.52 ± 0.04 | 645 ± 68 | -243 ± 22 | 40 ± 20 | -19 ± 30 | 340 | silicate |
| AH-147c | 4.41 ± 0.21 | 2.67 ± 0.05 | 151 ± 55 | 334 ± 26 | 10 ± 19 | -40 ± 20 | 375 | silicate |
| AH-148a | 13.40 ± 0.12 | 1.63 ± 0.01 | 2499 ± 30 | -186 ± 6 | 35 ± 8 | 59 ± 10 | 580 | silicate |
| AH-148c[#] | 8.66 ± 0.50 | 1.89 ± 0.10 | 1260 ± 131 | -59 ± 50 | -11 ± 53 | -119 ± 53 | 320 | silicate |
| AH-149 | 5.21 ± 0.38 | 1.45 ± 0.07 | 360 ± 100 | -279 ± 33 | -36 ± 55 | 81 ± 80 | 265 | silicate |
| AH-155a | 7.67 ± 0.30 | 1.95 ± 0.05 | 1003 ± 78 | -29 ± 24 | 82 ±42 | -2 ± 54 | 365 | silicate |
| AH-155b | 12.17 ± 1.05 | 1.90 ± 0.08 | 2176 ± 275 | -55 ± 38 | -12 ± 48 | 124 ± 70 | 290 | silicate |
| AH-157 | 7.52 ± 0.64 | 1.41 ± 0.08 | 963 ± 166 | -299 ± 38 | 50 ± 60 | -116 ± 85 | 290 | silicate |
| AH-161 | 5.92 ± 0.60 | 1.28 ± 0.09 | 546 ± 158 | -362 ± 45 | | | 305 | |
| AH-164 | 5.87 ± 0.38 | 1.97 ± 0.07 | 532 ± 98 | -16 ± 35 | 50 ± 54 | 12 ± 69 | 265 | silicate |
| AH-166a[%] | 6.43 ± 0.56 | 1.57 ± 0.06 | 680 ± 145 | -218 ± 31 | 36 ± 56 | 24 ± 68 | 305 | silicate |
| AH-166b | 7.02 ± 0.51 | 2.00 ± 0.08 | 832 ± 133 | -3 ± 40 | 60 ± 72 | -27 ± 114 | 265 | silicate |
| AH-167a | 7.43 ± 0.58 | 1.84 ± 0.08 | 941 ± 152 | -84 ± 38 | 66 ± 79 | -100 ± 82 | 290 | silicate |
| AH-167b | 3.91 ± 0.31 | 2.92 ± 0.08 | 22 ± 81 | 456 ± 42 | 31 ± 56 | 77 ± 70 | 330 | silicate |
| AH-38 | | | | | -281 ± 54 | -387 ± 61 | 175 | SiC X |
| AH-117b | | | | | -358 ± 50 | -202 ± 72 | 305 | SiC X |
| AH-99 | | | | | -68 ± 22 | 113 ± 34 | 265 | SiC Z |

Notes.— The grain phases are determined by NanoSIMS $Si^-/O^-$ and $MgO^-/O^-$ ratios. Grain sizes are determined from the data analysis software. Isotopic ratio errors are 1σ. Ratios given as δ-values are in permil (‰).

[%] Cross-sections of these grains were prepared and analyzed by TEM.
[&] Auger analyses of these grains indicate the phases deduced from NanoSIMS measurements were incorrect.
[#] Only C signal from contamination was detected in Auger spectroscopic analysis of these grains.

◊ Auger elemental maps were acquired for these grains and their surrounding regions, but individual grain spectra were not obtained. All of these grains are Mg-rich silicates except for AH-137b, which is Fe-rich.

Table 2. Oxygen, silicon, and carbon isotopic compositions, diameters, and phases of presolar grains identified in QUE 99177.

| Grain | $^{17}O/^{16}O$ ($\times10^{-4}$) | $^{18}O/^{16}O$ ($\times10^{-3}$) | $\delta^{17}O/^{16}O$ | $\delta^{18}O/^{16}O$ | $\delta^{29}Si/^{28}Si$ | $\delta^{30}Si/^{28}Si$ | $\delta^{13}C/^{12}C$ | Size (nm) | Phase |
|---|---|---|---|---|---|---|---|---|---|
| QUE99-20 | 6.09 ± 0.54 | 1.86 ± 0.08 | 591 ± 141 | -74 ± 41 | | | | 250 | oxide |
| QUE99-24 | 6.08 ± 0.26 | 1.52 ± 0.04 | 589 ± 67 | -240 ± 20 | | | | 415 | oxide |
| QUE99-28 | 5.34 ± 0.20 | 1.83 ± 0.04 | 396 ± 52 | -86 ± 19 | | | | 330 | oxide |
| QUE99-29 | 5.76 ± 0.15 | 1.76 ± 0.03 | 504 ± 38 | -120 ± 13 | | | | 350 | oxide |
| QUE99-38 | 4.10 ± 0.29 | 2.38 ± 0.08 | 71 ± 77 | 188 ± 13 | | | | 320 | oxide |
| QUE99-1 | 5.01 ± 0.24 | 2.00 ± 0.05 | 309 ± 61 | -5 ± 24 | | 11 ± 50 | | 290 | silicate |
| QUE99-2 | 4.94 ± 0.25 | 1.99 ± 0.04 | 290 ± 65 | -7 ± 22 | | -80 ± 44 | | 280 | silicate |
| QUE99-3 | 6.58 ± 0.28 | 2.07 ± 0.04 | 718 ± 72 | 32 ± 20 | | -35 ± 51 | | 320 | silicate |
| QUE99-4 | 5.16 ± 0.17 | 2.08 ± 0.03 | 347 ± 45 | 39 ± 17 | | 63 ± 46 | | 320 | silicate |
| QUE99-5 | 6.63 ± 0.33 | 2.08 ± 0.06 | 732 ± 85 | 37 ± 32 | | 50 ± 73 | | 250 | silicate |
| QUE99-6 | 6.48 ± 0.38 | 2.08 ± 0.07 | 693 ± 98 | 37 ± 34 | | 0 ± 76 | | 340 | silicate |
| QUE99-7 | 4.79 ± 0.15 | 1.98 ± 0.03 | 251 ± 39 | -12 ± 17 | | 29 ± 33 | | 340 | silicate |
| QUE99-8 | 4.88 ± 0.26 | 2.01 ± 0.04 | 274 ± 68 | 5 ± 21 | | 30 ± 50 | | 250 | silicate |
| QUE99-9 | 5.63 ± 0.26 | 1.97 ± 0.05 | 470 ± 67 | -16 ± 24 | | 66 ± 56 | | 290 | silicate |
| QUE99-10 | 5.73 ± 0.30 | 1.95 ± 0.06 | 495 ± 78 | -25 ± 27 | | -10 ± 59 | | 305 | silicate |
| QUE99-11 | 11.86 ± 0.47 | 1.85 ± 0.06 | 2099 ± 123 | -77 ± 28 | | 75 ± 60 | | 375 | silicate |
| QUE99-12 | 4.84 ± 0.20 | 2.07 ± 0.04 | 265 ± 51 | 32 ± 20 | | -54 ± 51 | | 280 | silicate |
| QUE99-13 | 5.03 ± 0.22 | 1.96 ± 0.04 | 313 ± 57 | -20 ± 21 | | 40 ± 47 | | 305 | silicate |
| QUE99-14 | 6.49 ± 0.34 | 1.95 ± 0.07 | 695 ± 88 | -28 ± 35 | -9 ± 8 | 13 ± 8 | | 340 | silicate |
| QUE99-15 | 5.15 ± 0.19 | 2.05 ± 0.03 | 346 ± 49 | 20 ± 17 | | -66 ± 40 | | 340 | silicate |
| QUE99-16 | 6.33 ± 0.23 | 1.91 ± 0.04 | 653 ± 59 | -48 ± 22 | | 59 ± 51 | | 305 | silicate |
| QUE99-17 | 5.19 ± 0.20 | 2.07 ± 0.04 | 355 ± 51 | 33 ± 21 | | 34 ± 55 | | 290 | silicate |
| QUE99-18 | 5.31 ± 0.23 | 2.00 ± 0.04 | 386 ± 60 | -1 ± 22 | | -4 ± 51 | | 320 | silicate |
| QUE99-19 | 5.89 ± 0.12 | 1.98 ± 0.02 | 539 ± 30 | -11 ± 10 | | -32 ± 34 | | 570 | silicate |
| QUE99-21 | 5.08 ± 0.27 | 1.53 ± 0.05 | 327 ± 71 | -239 ± 24 | | -4 ± 69 | | 265 | silicate |
| QUE99-22 | 5.30 ± 0.27 | 1.95 ± 0.06 | 383 ± 70 | -25 ± 28 | | -109 ± 55 | | 280 | silicate |
| QUE99-23 | 11.60 ± 0.32 | 1.90 ± 0.04 | 2028 ± 83 | -54 ± 21 | 146 ± 16 | 134 ± 22 | | 475 | silicate |
| QUE99-25 | 5.74 ± 0.27 | 1.53 ± 0.05 | 498 ± 70 | -237 ± 23 | | -46 ± 64 | | 355 | silicate |
| QUE99-26 | 6.13 ± 0.32 | 1.49 ± 0.05 | 600 ± 85 | -255 ± 25 | | -75 ± 58 | | 280 | silicate |
| QUE99-27 | 7.49 ± 0.34 | 1.30 ± 0.04 | 957 ± 88 | -352 ± 21 | | 10 ± 64 | | 385 | silicate |
| QUE99-30 | 4.99 ± 0.19 | 1.85 ± 0.04 | 302 ± 49 | -79 ± 21 | | 73 ± 41 | | 305 | silicate |
| QUE99-31 | 5.74 ± 0.21 | 1.63 ± 0.04 | 499 ± 55 | -188 ± 20 | | -68 ± 49 | | 280 | silicate |
| QUE99-32 | 6.91 ± 0.26 | 1.37 ± 0.04 | 805 ± 67 | -319 ± 21 | | 0 ± 39 | | 395 | silicate |
| QUE99-33 | 7.52 ± 0.41 | 1.23 ± 0.07 | 964 ± 106 | -386 ± 33 | | 196 ± 89 | | 405 | silicate |
| QUE99-34 | 5.98 ± 0.26 | 1.57 ± 0.05 | 561 ± 68 | -219 ± 25 | | -26 ± 59 | | 305 | silicate |
| QUE99-35 | 4.92 ± 0.31 | 1.65 ± 0.05 | 285 ± 80 | -176 ± 27 | | 42 ± 70 | | 265 | silicate |
| QUE99-36 | 3.57 ± 0.15 | 2.27 ± 0.04 | -68 ± 40 | 130 ± 19 | | 69 ± 44 | | 320 | silicate |
| QUE99-37 | 4.64 ± 0.28 | 3.05 ± 0.07 | 211 ± 73 | 520 ± 37 | | -11 ± 50 | | 355 | silicate |
| QUE99-39 | 5.84 ± 0.15 | 3.28 ± 0.04 | 525 ± 39 | 635 ± 21 | 1 ± 25 | -34 ± 29 | | 385 | silicate |
| QUE99-40 | | | | | | -44 ± 73 | 15000 ± 3000 | 365 | SiC B |
| QUE99-41 | | | | | | 0 ± 24 | 112 ± 38 | 265 | SiC |
| QUE99-42 | | | | | | -11 ± 45 | 437 ± 86 | 290 | SiC |
| QUE99-43 | | | | | -5 ± 39 | 42 ± 34 | 4800 ± 130 | 330 | SiC-M |
| QUE99-44 | | | | | | 29 ± 40 | 868 ± 75 | 330 | SiC |
| QUE99-45 | | | | | 136 ± 20 | 104 ± 18 | 1167 ± 47 | 305 | SiC-M |
| QUE99-46 | | | | | 34 ± 38 | 52 ± 20 | 486 ± 43 | 330 | SiC-M |
| QUE99-47 | | | | | | 0 ± 31 | 519 ± 72 | 320 | SiC |
| QUE99-48 | | | | | | 77 ± 32 | 799 ± 54 | 340 | SiC |
| QUE99-49 | | | | | | 43 ± 43 | 750 ± 130 | 290 | SiC |
| QUE99-50 | | | | | 67 ± 44 | 74 ± 21 | 571 ± 40 | 280 | SiC-M |
| QUE99-51 | | | | | | 76 ± 54 | 548 ± 93 | 340 | SiC |
| QUE99-52 | | | | | -87 ± 16 | 45 ± 16 | 985 ± 44 | 340 | SiC-Z |
| QUE99-53 | | | | | | -318 ± 27 | 78 ± 91 | 405 | $Si_3N_4$ |

Notes.—The phases are determined by NanoSIMS $Si^-/O^-$ and $Si^-/C^-$ ratios. For the five SiC grains analyzed for $\delta^{29}Si$, the reported $\delta^{30}Si$ values are weighted means of two measurements. Grains listed as "SiC" could either be mainstream (M) or type Z grains. The distinction cannot be made based on the existing isotopic data. The diameters are taken from manually defined areas in the data analysis software. Isotopic ratio errors are 1σ. Ratios given as δ-values are in permil (‰).

Table 3. Major element concentrations in atom% and phases of presolar grains in ALHA 77307.

| Grain | O | Si | Mg | Fe | Ni | Ca | Al | Mg/Si | Fe/Si | Mg # | Phase |
|---|---|---|---|---|---|---|---|---|---|---|---|
| AH-6[&] | 64 | | 9 | 7 | | | 20 | | | | oxide |
| AH-29[&,*] | 59 | | 6 | 14 | 5 | | 16 | | | | oxide |
| AH-33a | 56 | 25 | 7 | 12 | | | | 0.26 | 0.49 | 35 | silica |
| AH-9[&,#] | 55 | 14 | 12 | 19 | | | | 0.84 | 1.01 | 39 | Fe-rich silicate |
| AH-22[&] | 57 | 14 | 24 | 5 | | | | 1.67 | 0.34 | 83 | Mg-rich silicate |
| AH-23[&] | 57 | 14 | 21 | 8 | | | | 1.55 | 0.55 | 74 | Mg-rich silicate |
| AH-36 | 60 | 18 | 5 | 17 | | | | 0.29 | 0.95 | 23 | Fe-rich silicate |
| AH-42 | 56 | 21 | 18 | 5 | | | | 0.84 | 0.25 | 77 | Mg-rich silicate |
| AH-43 | 62 | 17 | 9 | 12 | | | | 0.52 | 0.74 | 41 | silicate |
| AH-46 | 63 | 16 | 10 | 11 | | | | 0.61 | 0.70 | 47 | silicate |
| AH-47 | 60 | 21 | 9 | 10 | | | | 0.44 | 0.46 | 49 | silicate |
| AH-48a | 56 | 11 | 15 | 17 | | | | 1.36 | 1.54 | 47 | silicate |
| AH-48b | 58 | 12 | 16 | 13 | | 2 | | 1.33 | 1.13 | 54 | silicate |
| AH-49a | 57 | 16 | 10 | 17 | | | | 0.61 | 1.04 | 37 | Fe-rich silicate |
| AH-50a[#] | 58 | 14 | 10 | 8 | | 4 | 6 | 0.71 | 0.62 | 54 | silicate |
| AH-50b[#] | 57 | 15 | 13 | 15 | | | | 0.84 | 1.0 | 46 | silicate |
| AH-56 | 62 | 16 | 11 | 11 | | | | 0.68 | 0.70 | 50 | silicate |
| AH-59 | 65 | 13 | 11 | 12 | | | | 0.87 | 0.91 | 49 | silicate |
| AH-60 | 60 | 20 | 8 | 12 | | | | 0.40 | 0.63 | 39 | Fe-rich silicate |
| AH-63 | 63 | 21 | 7 | 9 | | | | 0.33 | 0.41 | 44 | silicate |
| AH-65a | 55 | 21 | 8 | 4 | | 7 | 7 | 0.37 | 0.19 | 66 | Mg-rich silicate |
| AH-65b | 56 | 13 | 11 | 20 | | | | 0.85 | 1.61 | 35 | Fe-rich silicate |
| AH-72 | 55 | 17 | 11 | 18 | | | | 0.63 | 1.05 | 38 | Fe-rich silicate |
| AH-73a | 63 | 17 | 6 | 12 | | 2 | | 0.37 | 0.66 | 36 | Fe-rich silicate |
| AH-73b | 62 | 13 | 13 | 10 | | 2 | | 1.01 | 0.75 | 57 | silicate |
| AH-73c | 55 | 13 | 17 | 16 | | | | 1.35 | 1.28 | 51 | silicate |
| AH-75 | 58 | 17 | 12 | 13 | | | | 0.67 | 0.74 | 48 | silicate |
| AH-83a | 62 | 18 | 12 | 9 | | | | 0.65 | 0.47 | 58 | silicate |
| AH-83b | 61 | 17 | 7 | 8 | | | 8 | 0.43 | 0.46 | 48 | silicate |

| | | | | | | | | | | |
|---|---|---|---|---|---|---|---|---|---|---|
| AH-85a | 58 | 18 | 9 | 11 | 4 | | 0.49 | 0.61 | 45 | silicate |
| AH-85b | 53 | 17 | 19 | 11 | | | 1.10 | 0.61 | 64 | Mg-rich silicate |
| AH-85c | 56 | 18 | 15 | 11 | | | 0.81 | 0.63 | 56 | silicate |
| AH-85d | 55 | 13 | 16 | 11 | | 5 | 1.20 | 0.85 | 59 | silicate |
| AH-86 | 57 | 15 | 19 | 9 | | | 1.32 | 0.61 | 69 | Mg-rich silicate |
| AH-87a | 58 | 13 | 15 | 14 | | | 1.11 | 1.08 | 51 | silicate |
| AH-87b | 59 | 12 | 16 | 12 | | | 1.37 | 1.04 | 57 | silicate |
| AH-88a | 57 | 13 | 17 | 13 | | | 1.26 | 0.98 | 56 | silicate |
| AH-88b | 62 | 25 | 6 | 7 | | | 0.25 | 0.26 | 49 | silicate |
| AH-90[#] | 60 | 15 | 9 | 16 | | | 0.63 | 1.05 | 38 | Fe-rich silicate |
| AH-98a | 61 | 17 | 8 | 14 | | | 0.47 | 0.80 | 37 | Fe-rich silicate |
| AH-130 | 62 | 20 | 7 | 11 | | | 0.37 | 0.56 | 40 | silicate |
| AH-138a | 56 | 19 | 14 | 8 | 3 | | 0.72 | 0.43 | 62 | Mg-rich silicate |
| AH-138b | 57 | 19 | 15 | 10 | | | 0.83 | 0.51 | 62 | Mg-rich silicate |
| AH-138c | 57 | 16 | 8 | 12 | | 7 | 0.51 | 0.74 | 41 | silicate |
| AH-138d | 54 | 15 | 21 | 9 | | | 1.40 | 0.62 | 69 | Mg-rich silicate |
| AH-139b | 59 | 19 | 12 | 11 | | | 0.63 | 0.57 | 52 | silicate |
| AH-139c | 59 | 14 | 26 | | | | 1.83 | 0 | 100 | Mg-rich silicate |
| AH-140 | 56 | 17 | 12 | 14 | | | 0.71 | 0.83 | 46 | silicate |
| AH-142 | 60 | 19 | 11 | 10 | 2 | | 0.58 | 0.51 | 53 | silicate |
| AH-149 | 56 | 14 | 18 | 13 | | | 1.23 | 0.90 | 58 | silicate |
| AH-153 | 60 | 19 | 6 | 16 | | | 0.29 | 0.82 | 26 | Fe-rich silicate |
| AH-155a | 60 | 15 | 13 | 13 | | | 0.85 | 0.89 | 49 | silicate |
| AH-155b | 54 | 19 | 8 | 19 | | | 0.43 | 0.99 | 30 | Fe-rich silicate |
| AH-164 | 56 | 17 | 10 | 14 | 3 | | 0.62 | 0.83 | 43 | silicate |
| AH-166a | 56 | 17 | 8 | 4 | 6 | 8 | 0.48 | 0.21 | 69 | Mg-rich silicate |
| AH-166b | 55 | 15 | 10 | 21 | | | 0.63 | 1.37 | 32 | Fe-rich silicate |
| AH-167a | 53 | 18 | 15 | 15 | | | 0.81 | 0.86 | 49 | silicate |
| AH-167b | 51 | 18 | 17 | 14 | | | 0.93 | 0.76 | 55 | silicate |
| AH-144[%] | 70 | | | 30 | | | | | | Fe-rich oxide |
| AH-147b[%] | 54 | | | 10 | | 36 | | | | Al oxide |
| AH-132a[%] | 47 | 19 | 23 | 11 | | | 1.23 | 0.56 | 69 | Mg-rich silicate |
| AH-132b[%] | 55 | 22 | 14 | 10 | | | 0.63 | 0.47 | 58 | silicate |

| | | | | | | | | | | | |
|---|---|---|---|---|---|---|---|---|---|---|---|
| AH-133[%] | 54 | 10 | 24 | 12 | | | | 2.49 | 1.17 | 68 | Mg-rich silicate |
| AH-136[%] | 56 | 11 | 19 | 14 | | | | 1.80 | 1.30 | 58 | silicate |
| AH-139a[%] | 46 | 25 | 30 | | | | | 1.21 | 0 | 100 | Mg-rich silicate |
| AH-145[%] | 53 | 20 | 9 | 17 | | | | 0.47 | 0.87 | 35 | Fe-rich silicate |
| AH-147a[%] | 46 | 15 | 19 | 19 | | | | 1.27 | 1.24 | 51 | silicate |
| AH-147c[%] | 45 | 22 | 16 | 18 | | | | 0.71 | 0.82 | 47 | silicate |
| AH-148a[%] | 54 | 11 | 30 | 5 | | | | 2.72 | 0.47 | 85 | Mg-rich silicate |
| AH-148b[%] | 54 | 21 | 10 | 15 | | | | 0.47 | 0.72 | 40 | silicate |
| AH-148d[%] | 47 | 21 | 20 | 12 | | | | 0.96 | 0.56 | 63 | Mg-rich silicate |
| AH-151[%] | 52 | 20 | 19 | 10 | | | | 0.94 | 0.50 | 65 | Mg-rich silicate |
| AH-157[%] | 53 | 16 | 22 | 9 | | | | 1.41 | 0.57 | 71 | Mg-rich silicate |
| AH-161[%] | 55 | 18 | 24 | | | 2 | | 1.36 | 0 | 100 | Mg-rich silicate |

Notes.—The errors in the sensitivity factors are 3.6% for O, 11% for Si, 9.4% for Mg, 11.2% for Fe, 10.8% for Ca, and 24.9% for Al. These errors are based on the relative uncertainties in the sensitivity factors on silicate standards. Elemental concentration errors are 1-2 atom%. Mg# = Mg/(Mg+Fe)×100. Grains listed as "silicate" have comparable concentrations of Fe and Mg (Mg # between 40 and 60).

[&] These grains were initially identified by Nguyen et al. (2007b).
[*] The spectrum for this grain shows a clear Ni peak, but the actual abundance is uncertain due to the lack of a matrix-relevant sensitivity factor for this element.
[#] These grains experienced moderate charging during measurement, causing the elemental peaks to shift. Errors for these measurements are twice as large as for other grains.
[%] These spectra exhibit C peaks that are comparable to or larger than the O peak. The abundance determinations and detection limits are thus highly uncertain in these cases and should only be taken as rough estimates.

Table 4. Elemental concentrations in atom% of three presolar silicate grains as determined by STEM-EDX compared to Auger spectroscopy results.

| Grain | Analysis | O | Si | Mg | Fe | Ni | Ca | Al | Cr | Mg/Si | Fe/Si | Mg # |
|---|---|---|---|---|---|---|---|---|---|---|---|---|
| AH-139a | Auger | 46 | 25 | 30 | | | | | | 1.21 | 0 | 100 |
| | STEM whole grain | 46.73 (1.62) | 18.74 (0.17) | 24.28 (0.24) | 7.43 (0.07) | 0.75 (0.03) | 0.10 (0.02) | 1.79 (0.07) | 0.07 (0.02) | 1.30 | 0.40 | 77 |
| | STEM Point 1 | 42.35 (1.09) | 24.57 (0.16) | 22.20 (0.18) | 7.93 (0.06) | 0.78 (0.02) | 0.07 (0.01) | 1.95 (0.05) | 0.04 (0.01) | 0.90 | 0.32 | 74 |
| | STEM Point 2 | 56.4 (0.82) | 18.15 (0.13) | 18.62 (0.15) | 5.78 (0.05) | 0.14 (0.01) | 0.05 (0.01) | 0.81 (0.03) | 0.06 (0.01) | 1.03 | 0.32 | 76 |
| AH-166a | Auger | 56 | 17 | 8 | 4 | | 6 | 8 | | 0.48 | 0.21 | 69 |
| | STEM whole grain | 59.8 (1.64) | 15.8 (0.14) | 8.54 (0.13) | 4.21 (0.05) | 0.38 (0.02) | 3.51 (0.05) | 7.43 (0.11) | 0.28 (0.02) | 0.54 | 0.27 | 67 |
| AH-65a | Auger | 55 | 21 | 8 | 4 | | 7 | 7 | | 0.37 | 0.19 | 66 |
| | STEM whole grain | 61.05 (1.58) | 19.9 (0.19) | 5.97 (0.12) | 5.47 (0.07) | 1.16 (0.04) | 2.05 (0.05) | 4.22 (0.09) | 0.18 (0.02) | 0.30 | 0.27 | 52 |
| | STEM Point 1 | 63.62 (1.24) | 25.84 (0.20) | 2.51 (0.07) | 2.8 (0.05) | 0.95 (0.03) | 1.33 (0.04) | 2.89 (0.07) | ND | 0.10 | 0.11 | 47 |
| | STEM Point 2 | 60.62 (2.30) | 18.97 (0.19) | 4.73 (0.11) | 7.87 (0.09) | 1.91 (0.05) | 1.35 (0.04) | 4.39 (0.10) | 0.16 (0.02) | 0.25 | 0.41 | 38 |

Notes.—Concentration errors for the STEM-EDX data, given in parentheses, are reported as 2σ values derived from counting statistics and do not include possible k-factor variations for nanoscale glassy materials. The Auger spectrum of grain AH-139a suffered from significant C contamination and the elemental abundances are thus highly uncertain.